\documentclass[acmsmall, natbib=false, nonacm]{acmart}

\usepackage{libertine}
\usepackage[utf8]{inputenc}
\usepackage{siunitx}
\usepackage{nth}

\usepackage{centernot}

\usepackage[nameinlink]{cleveref}

\usepackage{booktabs}
\usepackage{tabularx}
\usepackage{multirow}
\usepackage{nicematrix}
\usepackage{tabularray}
\UseTblrLibrary{booktabs}

\usepackage[edges]{forest}

\usepackage{subcaption}
\usepackage{tikz}
\usepackage{pgfplots}
\usetikzlibrary{automata, positioning, arrows, arrows.meta, decorations.pathreplacing, calc, chains, shapes.misc, backgrounds, fit, perspective, matrix, shapes, shadows.blur, intersections, pgfplots.fillbetween, fillbetween, patterns, trees}

\pgfplotsset{
  compat=1.18,
  every axis plot/.append style={very thick},
  ecdf axis/.style={
    width=\textwidth,
    height=0.66\textwidth,
    enlargelimits=false,
    clip=false,
    grid=both,
    ymax=1.0,
    ymin=0.0,
    ytick = {0, 0.1, 0.2, 0.3, 0.4, 0.5, 0.6, 0.7, 0.8, 0.9, 1.0},
    legend cell align={left},
    xtick align=outside,
    ytick align=outside,
    every tick/.style={black},
    axis x line*=bottom,
    axis y line*=left,
  }
}

\pgfdeclarelayer{background}
\pgfsetlayers{background, main}

\usepackage{xcolor}
\usepackage{colorblind}

\colorlet{e1_color}{blue}
\colorlet{e2_color}{yellow}
\colorlet{e3_color}{green}
\colorlet{e4_color}{red}

\colorlet{trivago_color}{red}
\colorlet{spotify_color}{green}
\colorlet{microsoft_color}{blue}
\colorlet{android_color}{yellow}
\colorlet{vscode_color}{violet}
\colorlet{react_color}{cyan}

\colorlet{bounded_color}{blue}
\colorlet{leftbounded_color}{green}
\colorlet{rightbounded_color}{yellow}
\colorlet{unbounded_color}{red}

\acmJournal{TOSEM}

\usepackage[style=numeric, sortcites, backend=biber]{biblatex}
\def\rotateclockwise#1{
  \newdimen\xrw
  \pgfextractx{\xrw}{#1}
  \newdimen\yrw
  \pgfextracty{\yrw}{#1}
  \pgfpoint{\yrw}{-\xrw}
}

\def\rotatecounterclockwise#1{
  \newdimen\xrcw
  \pgfextractx{\xrcw}{#1}
  \newdimen\yrcw
  \pgfextracty{\yrcw}{#1}
  \pgfpoint{-\yrcw}{\xrcw}
}

\def\outsidespacerpgfclockwise#1#2#3{
  \pgfpointscale{#3}{
    \rotateclockwise{
      \pgfpointnormalised{
        \pgfpointdiff{#1}{#2}}}}
}

\def\outsidespacerpgfcounterclockwise#1#2#3{
  \pgfpointscale{#3}{
    \rotatecounterclockwise{
      \pgfpointnormalised{
        \pgfpointdiff{#1}{#2}}}}
}

\def\outsidepgfclockwise#1#2#3{
  \pgfpointadd{#2}{\outsidespacerpgfclockwise{#1}{#2}{#3}}
}

\def\outsidepgfcounterclockwise#1#2#3{
  \pgfpointadd{#2}{\outsidespacerpgfcounterclockwise{#1}{#2}{#3}}
}

\def\outside#1#2#3{
  ($ (#2) ! #3 ! -90 : (#1) $)
}

\def\cornerpgf#1#2#3#4{
  \pgfextra{
    \pgfmathanglebetweenpoints{#2}{\outsidepgfcounterclockwise{#1}{#2}{#4}}
    \let\anglea\pgfmathresult
    \let\startangle\pgfmathresult

    \pgfmathanglebetweenpoints{#2}{\outsidepgfclockwise{#3}{#2}{#4}}
    \pgfmathparse{\pgfmathresult - \anglea}
    \pgfmathroundto{\pgfmathresult}
    \let\arcangle\pgfmathresult
    \ifthenelse{180=\arcangle \or 180<\arcangle}{
      \pgfmathparse{-360 + \arcangle}}{
      \pgfmathparse{\arcangle}}
    \let\deltaangle\pgfmathresult

    \newdimen\x
    \pgfextractx{\x}{\outsidepgfcounterclockwise{#1}{#2}{#4}}
    \newdimen\y
    \pgfextracty{\y}{\outsidepgfcounterclockwise{#1}{#2}{#4}}
  }
  -- (\x,\y) arc [start angle=\startangle, delta angle=\deltaangle, radius=#4]
}

\def\corner#1#2#3#4{
  \cornerpgf{\pgfpointanchor{#1}{center}}{\pgfpointanchor{#2}{center}}{\pgfpointanchor{#3}{center}}{#4}
}

\def\hedgem#1#2#3#4{
  
  \outside{#1}{#2}{#4}
  \pgfextra{
    \def\hgnodea{#1}
    \def\hgnodeb{#2}
  }
  foreach \c in {#3} {
    \corner{\hgnodea}{\hgnodeb}{\c}{#4}
    \pgfextra{
      \global\let\hgnodea\hgnodeb
      \global\let\hgnodeb\c
    }
  }
  \corner{\hgnodea}{\hgnodeb}{#1}{#4}
  \corner{\hgnodeb}{#1}{#2}{#4}
  -- cycle
}

\def\hedgeii#1#2#3{
  \hedgem{#1}{#2}{}{#3}
}

\begin{document}

\title{The Capability of Code Review as a Communication Network}

\author{Michael Dorner}
\email{michael.dorner@bth.se}
\orcid{0000-0001-8879-6450}
\affiliation{%
  \institution{Blekinge Institute of Technology}
  \city{Karlskrona}
  \country{Sweden}
  \postcode{31749}
}

\author{Daniel Mendez}
\orcid{0000-0003-0619-6027}
\affiliation{%
  \institution{Blekinge Institute of Technology}
  \city{Karlskrona}
  \country{Sweden}
}
\affiliation{%
  \institution{fortiss}
  \city{Munich}
  \country{Germany}
}
\email{daniel.mendez@bth.se}

\begin{abstract}
  \noindent\textbf{Background:} %
  Code review, a core practice in software engineering, has been widely studied as a collaborative process, with prior work suggesting it functions as a communication network. However, this theory remains untested, limiting its practical and theoretical significance.

  \noindent\textbf{Objective:} %
  This study aims to (1) formalize the theory of code review as a communication network explicit and (2) empirically test its validity by quantifying how widely and how quickly information can spread in code review.

  \noindent\textbf{Method:} %
  We replicate an \emph{in-silico} experiment simulating information diffusion—the spread of information among participants—under best-case conditions across three open-source (Android, Visual Studio Code, React) and three closed-source code review systems (Microsoft, Spotify, Trivago) each modeled as communication network. By measuring the number of reachable participants and the minimal topological and temporal distances, we quantify how widely and how quickly information can spread through code review.

  \noindent\textbf{Results:} %
  We demonstrate that code review can enable both wide and fast information diffusion, even at a large scale. However, this capacity varies: open-source code review spreads information faster, while closed-source review reaches more participants.

  \noindent\textbf{Conclusion:} %
  Our findings reinforce and refine the theory, highlighting implications for measuring collaboration, generalizing open-source studies, and the role of AI in shaping future code review.
\end{abstract}



\maketitle

\section{Introduction}
\label{sec:introduction}

Modern software systems are often too large, too complex, and evolve too fast for an individual developer to oversee all parts of the software and, thus, to understand all implications of a change. Therefore, most software projects rely on code review to foster informal and asynchronous discussions on changes and their impacts before they are merged into the code bases. During those discussions, participants exchange information about the proposed changes and their implications, forming a communication network that emerges through code review.

This perspective on code review as a communication network is supported by a broad body of prior qualitative research. As a core practice in collaborative software engineering, the communicative and collaborative nature of code review has been examined in various studies \cite{Bacchelli2013, Baum20161, Bosu2017, Sadowski2018, Cunha2021}. Our synthesis of this prior research revealed a consistent pattern: practitioners widely perceive code review as a communication medium for information exchange, effectively functioning as a communication network \cite{Dorner2024upperbound}. This recurring pattern provides a solid foundation for the emerging \emph{theory of code review as a communication network}, which conceptualizes code review as a process that enables participants to exchange information around a code change.

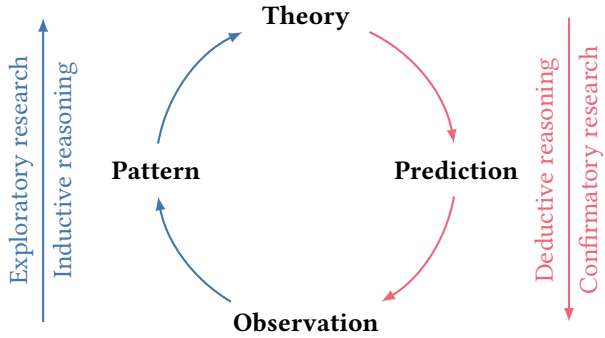
\begin{figure}
  \centering

  \begin{tikzpicture}
    \foreach \a/\t in {90/Theory, 0/Prediction, -90/Observation, -180/Pattern}{
      \node (\t) at (\a:2cm) {\textbf{\t}};
    }

    \draw[-latex, thick, red] (90-25:2cm) arc (90-25:0+10:2cm);
    \draw[-latex, thick, red] (0-10:2cm) arc (0-10:-90+30:2cm);
    \draw[-latex, thick, blue] (-90-30:2cm) arc (-90-30:-180+10:2cm);
    \draw[-latex, thick, blue] (-180-10:2cm) arc (-180-10:-270+25:2cm);

    \draw[latex-, thick, blue] (-3.5,2) -- (-3.5,-2) node[blue, midway, anchor=north, rotate=90] {Inductive reasoning} node[blue, midway, anchor=south, rotate=90] {Exploratory research};
    \draw[-latex, thick, red] (3.5,2) -- (3.5,-2) node[red, midway, anchor=south, rotate=90] {Deductive reasoning} node[red, midway, anchor=north, rotate=90] {Confirmatory research};
  \end{tikzpicture}
  \caption{The empirical research cycle (in analogy to \cite{Mendez2019}): While \textcolor{blue}{exploratory research} is theory-generating using inductive reasoning (starting with observations), \textcolor{red}{confirmatory research} is theory-testing using deductive reasoning (starting with a theory). This research is confirmatory.}
  \label{fig:inductivedeductivecycle}
\end{figure}

However, relying solely on this mostly exploratory and qualitative research to ground our understanding of code review as a communication network has shortcomings. Exploratory research begins with specific observations, distills patterns in those observations, typically in the form of hypotheses, and derives theories from the observed patterns through (inductive) reasoning. The nature of exploratory and especially qualitative research enables to analyze chosen cases and their contexts in great detail. However, their level of generalizability is also limited as they are drawn from those specific cases (especially when there is little to no relation to existing evidence that would allow for generalization by analogy). Therefore, while such initial theories may well serve as a very valuable foundation, they require still multiple confirmatory tests as, otherwise, their robustness (and scope) remains uncertain rendering it difficult to establish credibility, apply in practice, or develop it further. We thus postulate that exploratory (inductive) research alone is not sufficient to achieve a strong level of evidence, as discussed also in greater detail by \citeauthor{wohlin2013evidence}~\cite{wohlin2013evidence}. Confirmatory research, in turn, typically forms a set of predictions based on a theory (often in the form of hypotheses) and validates to which extent those predictions hold true or not against empirical observations (thus testing the consequences of a theory). To efficiently build a robust body of knowledge, we need both exploratory and confirmatory research to minimize bias and maximize the validity and reliability of our theories efficiently. \Cref{fig:inductivedeductivecycle} shows this interdependency between exploratory (theory-generating) and confirmatory (theory-testing) research in empirical research.

In this context, we have set out to address a gap in existing research by testing the theory of code review as a communication network. To that end, we focus on a phenomenon central to any communication network: its capability to spread information among its participants. We refer to this phenomenon as \emph{information diffusion}. From this perspective, we formulate our theory: code review, as a communication network, enables information to spread both widely and quickly among its participants. Whether this theory holds depends on the actual capabilities of code review systems to facilitate such information diffusion. If a given code review system exhibits little or no diffusion at all, it would challenge the general applicability of the theory to practical terms, suggesting that additional constraints, contextual factors, or boundary conditions must be considered to explain under which conditions and how code review functions effectively as a communication network.

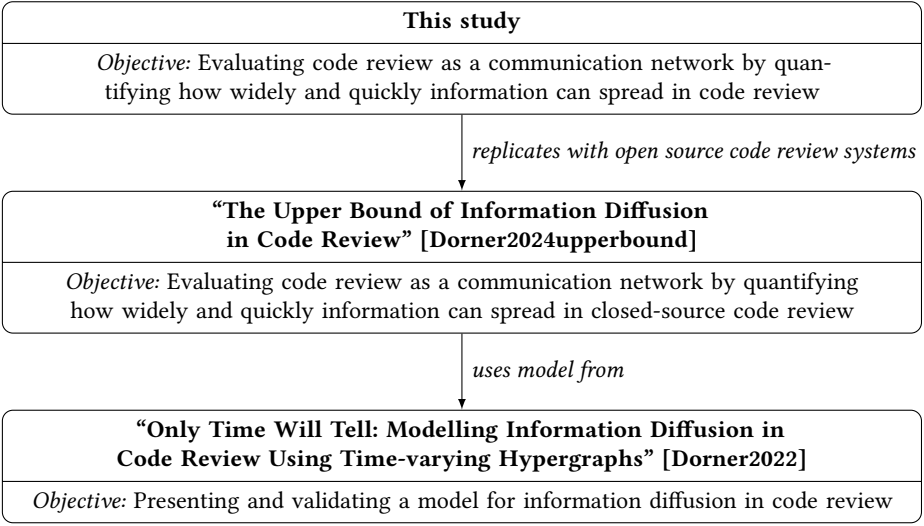
\begin{figure}
  \centering
  \begin{tikzpicture}[
      study/.style={rounded corners, rectangle split, rectangle split parts=2, text centered, minimum width=120mm, text width=120mm, font=\small, draw}
    ]
    \node[study] (esem) {\textbf{``Only Time Will Tell: Modelling Information Diffusion in Code Review Using Time-varying Hypergraphs'' \cite{Dorner2022}}
    \nodepart{second} \textit{Objective:} Presenting and validating a model for information diffusion in code review};
    \node[study, above=10mm of esem] (emse) {\textbf{``The Upper Bound of Information Diffusion in Code Review'' \cite{Dorner2024upperbound}}
    \nodepart{second} \textit{Objective:} Evaluating code review as a communication network by quantifying how widely and quickly information can spread in closed-source code review};
    \node[study, above=10mm of emse] (tosem) {\textbf{This study}
    \nodepart{second} \textit{Objective:} Evaluating code review as a communication network by quantifying how widely and quickly information can spread in code review};

    \draw[-latex] (emse) -- (esem) node[midway, right, font=\small\itshape] {uses model from} ;
    \draw[-latex] (tosem) -- (emse) node[midway, right, font=\small\itshape] {replicates with open source code review systems};

  \end{tikzpicture}
  \caption{The delineation and interplay with our prior work in this line of research.}
  \label{fig:delimitationinterplay}
\end{figure}

In prior work of this line of research (cf. \Cref{fig:delimitationinterplay}), we estimated the capabilities of closed-source code review by measuring how widely and how quickly\footnote{In this study, we use the terms \emph{widely} and \emph{quickly} instead of \emph{far} and \emph{fast}, as used in our prior study, to improve clarity and readability throughout---at the expense of a minor inconsistency between the two studies.} information can spread in code review systems at Microsoft, Spotify, and Trivago. Although the prior study as \emph{in-silico} experiment followed, in principle, confirmatory objectives, it did not explicitly formulate or test hypotheses and focused exclusively on closed-source code review systems---omitting a central phenomenon in software engineering: open-source software development. Open-source software development follow substantially different mechanics, including organizational structure \cite{Joblin2023} and developer liability and commitment \cite{Barcomb2020}, which may affect the communication within code review. We argue that studying and comparing both open-source and closed-source systems is essential for investigating the theory of code review as a communication network through a more holistic and context-sensitive lens.

The study at hands addresses those limitations and shall close our long-term investigations by answering the following two research questions considering both open-spource and closed-source software systems:

\begin{center}
  \begin{tabular}{@{}rl@{}}
    \textbf{RQ~1} & How widely can information spread within code review?\\
    \textbf{RQ~2} & How quickly can information spread within code review?\\[1em]
  \end{tabular}
\end{center}

In alignment with the baseline experiment \cite{Dorner2024upperbound}, we address the two research questions in an \emph{in-silico} experiment that simulates an artificial information diffusion within the code review within three open-source code review systems (Android, Visual Studio Code, and React) and three closed-source code review systems (Microsoft, Spotify, and Trivago), which we already used for the baseline experiment. The simulated information diffusion within the communication networks identifies all minimal time-respecting paths reflecting information diffusing through the communication network under best-case assumptions. The participants along those minimal time-respecting paths describe how wide information can spread among code review participants (RQ~1), and the minimal topological and temporal distances between participants describe how quickly information spreads (RQ~2). Together, both measures allow us to empirically test core assumptions of the theory of code review as communication network.

This work contributes to the understanding of code review as a communication network in software engineering in the following ways:

\begin{enumerate}
  \item \emph{Validating the theory of code review as communication network}\\
    We provide a large-scale and sophisticated simulation-based empirical validation that code review systems can support wide and fast information diffusion, reinforcing the theory of code review as a communication network.
  \item \emph{Comparative Analysis of Open-Source and Closed-Source Code Review Systems}\\
    Our study aims at critically analyzing and comparing different systems to yield a representative picture. This reveals systematic differences between open-source and closed-source code review environments: open-source systems enable faster diffusion, whereas closed-source systems support broader reach.
  \item \emph{Demonstrating the scalability of communication in large code review systems}\\
    We show that information diffusion in code review networks scales effectively, with no observed degradation in diffusion range or speed in larger systems, challenging assumptions that scale inherently limits communication efficiency.
  \item \emph{Analysis of scientific and practical implications} We analyze and highlight the key scientific and practical implications of the theory of code review as a communication network, now grounded in a broader and more robust empirical foundation.
  \item \emph{Comprehensive Replication Package}\\
    We release a complete, documented replication package including all datasets, thoroughly tested simulation code, and analysis scripts to support transparency, reproducibility, and future research by the community.
\end{enumerate}

In this manuscript, we borrow the definition of code review from the replicated study: We define code review as the informal and asynchronous discussion around a code change among humans. This means older results from formal code inspections and pair programming as an informal but synchronous discussion around a code change among usually two developers are beyond the scope of our study.

As a replication study, our work naturally shares substantial overlap with the original study we replicate \cite{Dorner2024upperbound}, including its definitions and the underlying model of information diffusion in code review \cite{Dorner2022}. \Cref{fig:delimitationinterplay} illustrates the delineation and interplay between the studies, situating our contribution within the broader line of research.
Too often, replication studies in software engineering research still tend to be ``not consistent in either the type of information reported or the level of detail reported'' \cite{Carver2010}. We therefore explicitly decided to provide the full explanations of the simulation model in \Cref{sec:simulation_model} and to add the inherently overlapping limitations \cite{Gomez2014} in \Cref{sec:threats_to_validity} to improve the readability and self-containing of this article.

\section{Background}
\label{sec:background}

In the following, we embed this study in the large picture on this line of research (\Cref{sec:line_of_research}), provide background information on replications in software engineering research (\Cref{sec:replications}), position \emph{in-silico} experiments as an empirical research method (\Cref{sec:insilicoexperiments}), and discuss prior work on closed-source and open-source code review (\Cref{sec:osvscs}). For a comprehensive discussion of the related work on measuring information diffusion in code review, we refer the reader to Section~2.2 in \citeauthor{Dorner2024upperbound}~\cite{Dorner2024upperbound}.

\subsection{Line of research}
\label{sec:line_of_research}

As already highlighted in \Cref{fig:delimitationinterplay}, this work forms part of a broader line of research. In the following, we highlight the two preceding studies and their contributions to this line of research.

In the first study of this line of research \cite{Dorner2022}, we presented and evaluated a model for information diffusion in code review. In an \emph{in-silico} experiment modelling the code review at Microsoft as our time-varying and as a traditional time-aggregated model, we found that traditional time-aggregated models significantly overestimate the paths of information diffusion available in communication networks and, thus, are neither precise nor accurate for modelling and measuring the spread of information within communication networks that emerge from code review. We will discuss the model as part of \Cref{sec:simulation_model} in detail.

In our prior work \cite{Dorner2024upperbound}, we synthesized the findings from prior qualitative work emerging in the theory of code review as communication networks. We then used the novel model from \cite{Dorner2022} to simulate an artificial information diffusion within large (Microsoft), mid-sized (Spotify), and small code review systems (Trivago). We measure the minimal topological and temporal distances between the participants to quantify how widely and how quickly information can spread in code review, which forms the upper bound of information diffusion in code review. This \emph{in-silico} experiment serves as the baseline experiment for our replication. As such, we discuss the details of the findings in detail as part of our \Cref{sec:discussion}. Despite its confirmatory nature and the research design as an experiment, we lost out on explicitly framing this study as a confirmatory study with explicit hypotheses. We remedy this shortcoming in this study.

\subsection{Replications in Software Engineering}
\label{sec:replications}

To consolidate a body of knowledge built upon evidence, experimental results have to be extensively verified. Replications can serve this purpose and verify the validity and reliability of previous findings. 

Although replications are a cornerstone in science and replication studies in software engineering research are generally gaining much attention in recent years~\cite{cruz2019replication}, there is no common interdisciplinary (let alone intradisciplinary) understanding of replications~\cite{GonzalezBarahona2023}\footnote{We deem to mention at this point that ACM swapped the terms ``reproducibility'' and ``replication'' in August 2020 to ``harmonize its terminology and definitions with those used in the broader scientific research community'' \cite{acmbadges}.}.

In our work, we follow the definitions and classification of \citeauthor{Gomez2014}. In their, to us inspiring work, the authors proposed a classification of replication types within software engineering research~\cite{Gomez2014}. They identified three different types of replications: literal, operational, and conceptual replications. Operational replications can be altered by four dimensions of experimental configurations: protocol, operationalizations, populations, and experimenters.

\begin{table}
  \caption{Replication types for experiments in software engineering (adapted from \cite{Gomez2014}). This study is an operational replication in which we varied the population (highlighted in \textbf{bold}).}
  \label{tab:replicationtypes}
  \begin{tabularx}{\textwidth}{@{}l l X@{}}
    \toprule
    Type & Varied dimension & Description \\
    \midrule
    Literal & None & Repetition. The aim is to run a replication of the baseline experiment as exactly as possible. The replication is run by the same experimenters using the same protocol and the same operationalizations on different samples of the same population. \\
    \textbf{Operational} & Protocol & The experimental protocol elements are varied with the aim of verifying that the observed results are reproduced using equivalent experimental protocols. \\
    & Operationalization & The cause and/or effect operationalizations are varied in order to verify the bounds of the cause and/or effect construct operationalizations within which the results hold. \\
    & \textbf{Population} & \textbf{The populations are varied to verify the limits of the populations used in the baseline experiment.} \\
    & Experimenter & The experimenters are varied to verify their influence on the results. \\
    Conceptual & Unknown & Reproduction. Different experimenters run the replication with new protocols and operationalizations. \\
    \bottomrule
  \end{tabularx}
\end{table}

Following this classification, our study is an operational replication where we varied the population to verify the limits of the population of code review systems used in the baseline experiment. \citeauthor{Gomez2014} suggests the name \emph{changed-populations replication} (highlighted in \Cref{tab:replicationtypes}) for our type of replication.

As already pointed out in \Cref{sec:introduction}, replication studies in software engineering still tend to be ``not consistent in either the type of information reported or the level of detail reported'' \cite{Carver2010}. \citeauthor{Carver2010} proposed a starting point for a discussion on reporting guidelines for experimental replications in software engineering research \cite{Carver2010}. For our study, we follow those reporting guidelines as we include the key aspects of the original study, report the motivation for conducting the replication (\Cref{sec:introduction}), discuss the threats of validity of the overlapping in experimenters in both experiments (\Cref{sec:threats_to_validity}), and we will compare the results of the original study as part of the discussion (\Cref{sec:discussion}) which serves as foundation for validating our theory.

\subsection{In-silico Experiments in Software Engineering}
\label{sec:insilicoexperiments}

In this \emph{in-silico} experiment, we simulate an artificial information diffusion and measure the resulting traces generated by the spread of the information \cite{Dorner2024upperbound}. Given the rarity of simulations in the empirical software engineering community, we motivate the explicit choice of that as our empirical research method.

Simulations are experiments with a model. Simulations play to their strengths when traditional experiments would have been too complex, too expensive, too lengthy, or simply not possible or accessible otherwise. Those attributes match the objectives and setting of our research since experiments with whole software companies or open-source components are not feasible.

If all parts of the experiment, i.e. subjects and settings, are modelled as software (see \Cref{tab:simulation_types}), the simulation becomes an \emph{in-silico} experiment.

\begin{table}
  \centering
  \caption{A comparison of \emph{in-vivo}, \emph{in-vitro}, \emph{in-virtuo}, and \emph{in-silico} experiments with respect to  subjects and settings.}
  \label{tab:simulation_types}
  \vspace{0.5cm}
  \begin{NiceTabular}{@{}lllll@{}}
    \CodeBefore [create-cell-nodes]
    \begin{tikzpicture}
      \node (n0) [draw, inner sep=2pt, rounded corners = 1mm, fit= (4-3)] {};
      \node (n1) [inner sep=2pt, fit=(3-5)] {};
      \node (n2) [inner sep=2pt, fit=(4-4)] {};
      \node (n3) [inner sep=2pt, fit=(4-5)] {};
      \draw[rounded corners = 1mm, fill=yellow!50] (n1.west) |- (n2.north) -| (n2.west) |- (n3.south) -| (n1.north east) -| cycle;
      \draw (n3.10) to[out=0, in=-90] (9,3) node[text width=2cm, text centered, anchor=south] {as computer\\(software) model};
      \draw (n0.20) to[out=90, in=-90] (5,3.5) node[anchor=south] {otherwise};
      \node (a1) [inner sep=2pt, fit=(4-2)] {};
      \node (a2) [inner sep=2pt, fit=(4-5)] {};
      \draw[thick, <->] ([yshift=-3em]a1.west) node[below, anchor=north west] {less control} -- ([yshift=-3em]a2.east) node[below, anchor=north east] {more control};
      \draw[thick, <->] ([yshift=-6em]a1.west) node[below, anchor=north west] {more realistic} -- ([yshift=-6em]a2.east) node[below, anchor=north east] {less realistic};
      \draw[thick, <->] ([yshift=-9em]a1.west) node[below, anchor=north west, text width=2cm, align=left] {implicit\\assumptions} -- ([yshift=-9em]a2.east) node[below, anchor=north east, text width=2cm, align=right] {explicit\\assumptions};
    \end{tikzpicture}
    \Body
    \toprule
    & \multicolumn{4}{l}{Experiment} \\
    \cmidrule(r){2-5}
    & {in-vivo} & {in-vitro} & {in-virtuo} & {in-silico} \\
    \midrule
    {Subjects} & natural & natural & natural & modelled \\
    {Settings} & natural & modelled & modelled & modelled \\
    \bottomrule
    \\[3cm]
  \end{NiceTabular}
\end{table}

Although simulations have been applied in different research fields of software engineering, e.g., process engineering, risk management, and quality assurance \cite{Muller2008}, the role of simulations as an empirical method is still often subject to some form of prejudice but also subject to ongoing more philosophical debates. \citeauthor{Stol2018}, for example, positioned computer simulations in their ABC framework \cite{Stol2018} in a non-empirical setting because, as the authors argue: ``while variables can be modelled and manipulated based on the rules that are defined within the computer simulation, the researcher does not make any new empirical observations of the behavior of outside actors in a real-world setting (whether these are human participants or systems)'' \cite{Stol2018}. Without discussing the role of simulations in the empirical software engineering community to the extent they might deserve, however, we still argue for their suitability as an evidence-based (empirical) approach in our context where observations would otherwise not be possible (or, at least, not realistic).

We consider computer simulations as an empirical research method, the same as done in other disciplines and inter-disciplines (where, for instance, climate simulations are the first-class citizens in the set of research methods). Empirical research methods are ``research approaches to gather observations and evidence from the real world''\cite{Stol2018} and same as in other empirical research methods, in simulation models, we build the models based on real-world observations and make conclusions based on the empirical observations along the execution (in our case, of the simulations). These simulation models are abstractions from the real world---same as the (often implicit) theoretical models underlying quasi-controlled (in-vitro) experiments. Simulations and their underlying models further abstract from (and make explicit) complex systems and make observations and evidence possible in situations where more traditional experiments are rendered infeasible (e.g., too expensive, dangerous, too long, or not accessible) or simply impossible at all; for instance, observations when exploring the capabilities of real-world communication networks with thousands of developers as done in the simulation study presented in the manuscript at hands.

Needless to elaborate, a certain abstraction from the real world is inherent to all empirical research methods, either in the form of explicit models or implicit assumptions. Like every measurement, the models we create come with certain accuracy and precision---with a certain quality. However, we may still argue that the quality of a research method does not necessarily decide upon whether it qualifies as empiricism or not but rather the underlying constructs and their (evidence-based) sources. To avoid surreal models and ensure the quality of a model, however, the modelling itself needs to be guided by quality assurance in verification and validation, and the sample used needs to be realistic; both would, in turn, be in tune with the underlying arguments by available positionings such as the one by \cite{Stol2018}. To increase the transparency in the quality of our simulations, we further disclose all developed software components as a replication package, also including the open-source communication networks we used as a sample.

\subsection{Open-Source vs. Closed-Source Code Review}
\label{sec:osvscs}

Insights and results from open source are not easily transferrable to closed-source software development in general since the definition does not refer to a specific software development process. Open-source software is software that is available under a license that grants the right to use, modify, and distribute the software---modified or not---to everyone free of charge \parencite[adapted from][]{Riehle2023}. This definition does not imply any specific software development or quality assurance process. However, the public availability of the source code alone is a necessary but still not sufficient condition.

The right to modify open-source code makes code review a cornerstone for quality assurance in open source: Open-source projects try to feed back modifications to benefit from improving or enhancing their projects and, thereby, create a larger community around the open-source project. However, unlike closed-source software development, where developers have an employment agreement, a developer contributing to an open-source project has no formal contract with the open-source project specifying conditions, obligations, and commitment. Contributors in open source may make contributions only once, for a short duration, or infrequently and irregularly \cite{Lee2017,Barcomb2019,Barcomb2020}. Open-source project maintainers may have to maintain the source code contributed by themselves or find a developer who can, which can be difficult without formal managerial authority over open-source developers. Therefore, the lack of formal managerial authority leaves code review the only way for open-source maintainers to exercise power over open-source developers. This underlying motivation makes any comparison of open-source and closed-source code review inherently challenging.

Two seminal studies accepted the challenge and compared code review in open-source and closed-source software development with respect to code review practices and quantitative measurements on the one hand and expectations towards code review on the other.

First, \citeauthor{Rigby2013} compared code review practices from open-source and closed-source software development efforts in order to distill code review practices \cite{Rigby2013}. The sample of code review systems ranges from closed-source software development projects at Microsoft (Bing, SQL Server, and Office 2013) and AMD, as well as open-source code review systems. The authors implicitly distinguish between community-led (i.e., Apache, Subversion, Linux, FreeBSD, KDE, and Gnome) and Google-led open-source projects (Android\footnote{Throughout this paper, we refer to the Android Open Source Project simply as \emph{Android}.} and Chromium OS\footnote{The open-source project that backs the operating system \emph{Chrome OS} is officially named \emph{Chromium OS}. We took the liberty to use the official name here.}). Although we explicitly exclude code review practices as a sampling dimension for our approach, narrowing down the large diversity within open source to company-led open-source projects allows us to focus on relevant and reasonably large open-source projects. Surprisingly, the study finds little difference between the different code review systems studied in the context of open-source or closed-source software development.

Second, \citeauthor{Bosu2017} reports the findings of an interview study with a well-designed survey of code review participants from open-source and closed-source software development, namely Microsoft,  \cite{Bosu2017}. They mined the code review systems of 34\footnote{The introduction speaks of 36 open-source projects.} open-source projects ``that used either Gerrit, ReviewBoard, or Rietveld to identify who had participated in at least 30 code review requests``. Unfortunately, we could not retrieve how the sampling frame \cite{Baltes2022} of open-source projects was selected or constructed.

Both studies found a large overlap between code review in the context of open-source and closed-source software development with respect to their respective research area. \citeauthor{Rigby2013} finds little difference between the different code review systems studied in the context of open-source or closed-source software development with respect to different measurements of code review systems in both worlds \cite{Rigby2013}. \cite{Bosu2017} also finds that ``the results were similar for the [open-source] and Microsoft developers.'' with the following two notable exceptions: The participants from open source reported building a relationship between the code review participants is the most important aspect while code review participants from Microsoft consider information diffusion (``knowledge dissemination'') more important. Similarly, open-source code review participants take the relationship with the code author and his or her reputation into account most prominently to decide whether to contribute to a code review. Conversely, for code review participants at Microsoft, the expertise of the code author (i.e., if a developer writes good code that s/he can learn) is the most important aspect.

In our manuscript, we will continue and extend this discussion on the equivalence of code review in closed-source and open-source software development to information diffusion within code review.

\section{Experimental Design}
\label{sec:experimental_design}

In this section, we describe the design of our \emph{in-silico} experiment that evaluates and quantifies how widely (RQ~1) and how quickly (RQ~2) information can diffuse in code review and postulate our hypotheses.

The underlying idea of our \emph{in-silico} experiment is, in principle, simple: We model different code review systems as communication networks and compute all minimal distances between all code review participants. The cardinality of reachable participants indicates how wide (RQ~1) information can spread, and distances between participants indicate how quickly (RQ~2) information can spread in code review. Since we used minimal paths and created the communication networks under best-case assumptions, the results describe the upper bound of information diffusion in code review \cite{Dorner2024upperbound}.

Yet communication—and, by extension, information diffusion—is (1) inherently time-dependent and (2) not strictly bilateral, as code reviews often involve exchanges among more than two participants. As a result, traditional graphs are not well suited for modeling such interactions and tend to dramatically overestimate information diffusion \cite{Dorner2022}. To address this, we use time-varying hypergraphs to model the communication network and to compute the shortest paths between all vertices. Hypergraphs generalize traditional graphs, allowing the use of standard algorithms such as Dijkstra’s for computing minimal-path distances. However, in time-varying hypergraphs, the distance between two vertices can be both topological (i.e., the fewest hops) and temporal (i.e., the shortest duration). This dual characterization enables us to answer RQ~2 by measuring both types of distance. Importantly, both topological and temporal distances yield the same set of reachable participants, which serves as the basis for answering RQ~1.

Since all models are per definition abstractions and, accordingly, simplifications of reality, the quality of an \emph{in-silico} experiment highly depends on the quality of the simulation model and its parametrization. Therefore, we provide a more elaborate description of our simulation model, which was originally proposed and partially validated in \cite{Dorner2022}, in \Cref{sec:simulation_model} and its parametrization of its computer model by empirical code review data (\Cref{sec:parametrization}). Before that, however, we first derive and explicitly state our hypotheses the theory consists of in greater detail.

\subsection{Hypotheses}

If code review is a functional communication network that enables the exchange of information (theory $T$) as identified by different exploratory studies \cite{Dorner2024upperbound}, then code review spreads information quickly (hypothesis $H_2$) and widely (hypothesis $H_1$) among its participants. We can formulate this sentence as the propositional statement

\begin{equation}
  T \implies \left(H_1 \land H_2\right).
\end{equation}

That means our theory $T$ can be falsified in its universality if we can reject at least one of our hypotheses encoding the theory of code review as communication network. Note that our ambition is not as much to achieve a binary answer to our hypothesis tests by falsifying a theory, but rather to gain further insights through the experimental, confirmatory process. In the absence of a predefined threshold for when to reject our hypotheses, we further adopt a more data-driven interpretation grounded in patterns observed across our measurements. That is, rather than establishing an arbitrary cutoff \emph{a priori}, we discuss the empirical upper bound of information diffusion within our sample of code review systems to identify consistent deviations from the expectations encoded in the hypotheses.

\subsection{Simulation Model}
\label{sec:simulation_model}

In general, a simulation model consists of two components: (1) the conceptual model describing our derivations and assumptions and (2) the computer model as the implementation of the conceptual model. The following two subsections describe each component in detail.

\subsubsection{Conceptual Model}
\label{sec:conceptual_model}

In the following, we describe how we conceptually model communication networks from code review discussions and the information diffusion within those communication networks.

\paragraph{Communication Network}
\label{sec:communication_network}

Communication, the purposeful, intentional, and active exchange of information among humans, does not happen in the void. It requires a channel to exchange information. A \emph{communication channel} is a conduit for exchanging information among communication participants. Those channels are

\begin{enumerate}
  \item {\itshape multiplexing}---A channel connects all communication participants sending and receiving information.
  \item {\itshape reciprocal}---The sender of information also receives information and the receiver also sends information. The information exchange converges. This can be in the form of feedback, queries, or acknowledgments. Pure broadcasting without any form of feedback does not satisfy our definition of communication.
  \item {\itshape concurrent}---Although a human can only feed into and consume from one channel at a time, multiple concurrent channels are usually used.
  \item {\itshape time-dependent}---Channels are not available all the time; after the information is transmitted, the channels are closed.
\end{enumerate}

Channels group and structure the information for the communication participants over time and content. Over time, the set of all communication channels forms a communication network among the communication participants.

In the context of our study on information diffusion, a communication channel is a discussion in a merge (or pull) request. A channel for a code review on a merge request begins with the initial submission and ends with the merge in case of an acceptance or a rejection. All participants of the review of the merge request feed information into the channel and, thereby, are connected through this channel and exchange information they communicate. After the code review is completed and the discussion has converged, the channel is closed and archived, and no new information becomes explicit and could emerge. However, a closed channel is usually not deleted but archived and is still available for passive information gathering. We do not intend to model this passive absorption of information from archived channels by retrospection with our model. For this line of research, we recommend the work by \cite{Pascarella2018} as further reading.

From the previous postulates on channel-based communication, we derive our mathematical model: Each communication medium forms an undirected, time-varying hypergraph in which hyperedges represent communication channels. Those hyperedges are available over time and make the hypergraph time-dependent. Additionally, we allow parallel hyperedges\footnote{This makes the hypergraph formally a \emph{multi-hypergraph} \cite{Ouvrard2020}. However, we consider the difference between a hypergraph and a multi-hypergraph as marginal since it is grounded in set theory. Sets do not allow multiple instances of the elements. Therefore, instead of a set of hyperedges, we use a multiset of hyperedges that allows multiple instances of the hyperedge.}---although unlikely, multiple parallel communication channels can emerge between the same participants at the same time but in different contexts.

Such an undirected, time-varying hypergraph reflects all four basic attributes of channel-based communication:

\begin{itemize}
  \item {\itshape multiplexing}---since a single hyperedge connects multiple vertices,
  \item {\itshape concurrent}---since (multi-)hypergraphs allow parallel hyperedges,
  \item {\itshape reciprocal}---since the hypergraph is undirected, information is exchanged in both directions, and
  \item {\itshape time-dependent}---since the hypergraph is time-varying.
\end{itemize}

In detail, we define the channel-based communication model for information diffusion in the timeframe $\mathcal{T}$ to be an undirected time-varying hypergraph
\[\mathcal{H} = (V, \mathcal{E}, \rho, \xi, \psi)\]
where

\begin{itemize}
  \item $V$ is the set of all human participants in the communication as vertices
  \item $\mathcal{E}$ is a multiset (parallel edges are permitted) of all communication channels as hyperedges,
  \item $\rho$ is the \emph{hyperedge presence function} indicating whether a communication channel is active at a given time,
  \item $\xi \colon E \times \mathcal{T} \rightarrow \mathbb{T}$, called \emph{latency function}, indicating the duration to exchange information among communication participants within a communication channel (hyperedge),
  \item $\psi \colon V \times \mathcal{T} \rightarrow \{0, 1\}$, called \emph{vertex presence function}, indicating whether a given vertex is available at a given time.
\end{itemize}

\paragraph{Information Diffusion}\label{sec:information_diffusion}

The time-respecting routes through the communication network are potential \emph{information diffusion}, the spread of information. To estimate the upper bound of information diffusion and, thereby, answer both of our research questions, we measure the distances between the participants under best-case assumptions.

For information diffusion in code review, we made the following assumptions:

\begin{itemize}
  \item {\itshape Channel-based}---Information can only be exchanged along the information channels that emerged from code review. The information exchange is considered to be completed when the channel is closed.
  \item {\itshape Perfect caching}---All code review participants can remember and cache all information in all code reviews they participate in within the considered time frame.
  \item {\itshape Perfect diffusion}---All participants instantly pass on information at any occasion in all available communication channels in code review.
  \item {\itshape Information diffusion only in code review}---For this simulation, we assume that information gained from discussions in code review diffuses only through code review.
  \item {\itshape Information availability}---To have a common starting point and make the results comparable, the information to be diffused in the network is already available to the participant, which is the origin of the information diffusion process.
\end{itemize}

Our assumptions make the results of the information diffusion a best-case scenario. Although the assumptions do not likely result in actual, real-world information diffusion, they serve well the scope of our study, namely to quantify the upper bound of information diffusion.

The possible routes through the communication network describe how information can spread through a communication network. Those routes are time-sensitive: a piece of information gained from a communication channel (i.e., a code review discussion) can be shared and exchanged in all subsequent communication channels but not in prior, closed communication channels.

Mathematically, those routes are time-respecting walks, so-called \emph{journeys}, in a time-varying hypergraph representing the communication network. A journey is a sequence of tuples
\[ \mathcal{J} = \left\{(e_1, t_1), (e_2, t_2), \dots, (e_k, t_k),\right\} \]
such that $\{e_1, e_2, \dots, e_k\}$ is a walk in $\mathcal{H}$ with $\rho(e_i, t_i) = 1$ and $t_{i+1} > t_i + \xi(e_i, t_i)$ for all $i < k$.

We define $\mathcal{J}^*_{\mathcal{H}}$ the set of all possible journeys in a time-varying graph $\mathcal{H}$ and $\mathcal{J}^*_{(u, v)} \in \mathcal{J}^*_{\mathcal{H}}$ the journeys between vertices $u$ and $v$. If $\mathcal{J}^*_{(u, v)} \neq \emptyset$, $u$ can reach $v$, or in short notation $u \leadsto v$.\footnote{In general, journeys are not symmetric and transitive---regardless of whether the hypergraph is directed or undirected: $u \leadsto v \centernot\Leftrightarrow v \leadsto u$.} Given a vertex $u$, the set $\{v \in V \colon u \leadsto v \}$ is called \emph{horizon} of vertex $u$.

The notion of length of a journey in time-varying hypergraphs is two-fold: Each journey has a topological distance (measured in number of hops) and temporal distance (measured in time). This gives rise to two distinct definitions of distance in a time-varying graph $\mathcal{H}$:

\begin{itemize}
  \item The \emph{topological distance} from a vertex $u$ to a vertex $v$ at time $t$ is defined by $d_{u, t}(v) = \min \{\vert \mathcal{J}(u, v) \vert_{h}\}$ where the journey length is $\vert \mathcal{J}(u, v) \vert_{h} = \vert \{e_1, e_2, \dots, e_k\} \vert$. This journey is the \emph{shortest}.
  \item The \emph{temporal distance} from a vertex $u$ to a vertex $v$ at time $t$ is defined by $\hat{d}_{u, t}(v) = \min \{\psi(e_k) + \xi(e_k) - \xi(e_1)\}$.\footnote{In our case, $\psi(e_k)$ is always $0$.} This journey is the \emph{fastest}.\footnote{For the interested reader, we would like to add that if the temporal distance is not defined for a relative time but for an absolute time $\hat{d}_{u, t}(v) = \min \{\psi(e_k) + \xi(e_k)\}$, the journey is called \emph{foremost}. For this line of research, the foremost journeys are not used.}
\end{itemize}

With this conceptual model and its mathematical background, we are now able to answer both research questions by measuring two characteristics of all possible routes through the communication network:
\begin{itemize}
  \item The distribution of the horizon of each participant in a communication network represents how wide information can spread (RQ~1).
  \item The distribution of all shortest and fastest journeys between all participants in a communication network answers how fast information can spread in code review (RQ~2). We measure how fast information can spread in code review in terms of the topological distance (minimal number of code reviews required to spread information between two code review participants) and the temporal distance (minimal timespan to spread information between two code review participants).
\end{itemize}

Those measurements within code review communication networks will result in the upper bound of information diffusion in code review.

\subsubsection{Computer Model}
\label{sec:computer_model}

Since our mathematical model is not trivial and lacks performant tool support for time-varying hypergraphs, we dedicate this section to the computer model and the implementation of the mathematical model described previously.

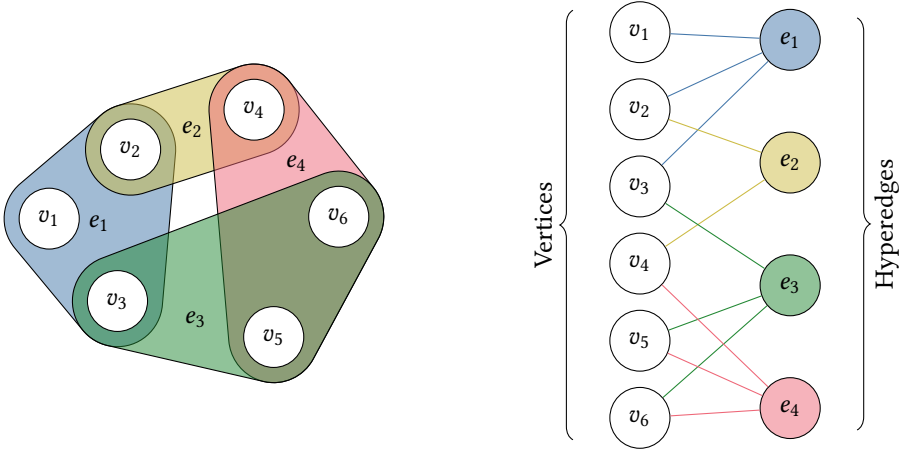
\begin{figure}
  \centering
  \begin{subfigure}[t]{0.48\textwidth}
    \captionsetup{skip=1em}%
    \centering
    \begin{tikzpicture}
      \tikzstyle{hedge}=[draw, circle, minimum size=8mm, fill=white];

      \begin{scope}[rotate=-5, xshift=0.0cm]
        \node at (1,1) (v1) {};
        \node at (2,2) (v2) {};
        \node at (2,0) (v3) {};

      \end{scope}

      \begin{scope}[rotate=5, yshift=0.0cm]
        \node at (4,2) (v4) {};
        \node at (4,-1) (v5) {};
        \node at (5,0.5) (v6) {};
      \end{scope}

      \node at (0,-2) {}; 

      \draw[draw, fill=e1_color, fill opacity=0.5] \hedgem{v1}{v2}{v3}{6mm};
      \draw[draw, fill=e2_color, fill opacity=0.5] \hedgeii{v2}{v4}{6mm};
      \draw[draw, fill=e4_color, fill opacity=0.5] \hedgem{v6}{v5}{v4}{6mm};
      \draw[draw, fill=e3_color, fill opacity=0.5] \hedgem{v6}{v5}{v3}{6mm};

      \node[hedge] at (v1) {$v_1$};
      \node[hedge] at (v2) {$v_2$};
      \node[hedge] at (v3) {$v_3$};
      \node[hedge] at (v4) {$v_4$};
      \node[hedge] at (v5) {$v_5$};
      \node[hedge] at (v6) {$v_6$};

      \node at (barycentric cs:v1=1,v2=1,v3=1) {$e_1$};
      \node at (barycentric cs:v2=1,v4=1) {$e_2$};
      \node at (barycentric cs:v3=1,v5=1) {$e_3$};
      \node at (barycentric cs:v4=1,v6=1) {$e_4$};

    \end{tikzpicture}
    \caption{An example time-varying hypergraph whose so-called hyperedges (denoted by $e_\square$) can link any arbitrary number of vertices (denoted by $v_\square$): For example, hyperedge $e_3$ connects three vertices. The horizon and minimal paths of vertex depend highly on the temporal order of the hyperedges: for example, the horizon of $v_1$ contains all vertices if the temporal availabilities of the hyperedges are $e_1 < e_2 < e_4 < e_3$, but none if $e_1 > e_2 \geq e_3$.}
    \label{fig:example_hypergraph}
  \end{subfigure}%
  \hfill%
  \begin{subfigure}[t]{0.48\textwidth}
    \captionsetup{skip=1em}%
    \centering
    \begin{tikzpicture}
      \tikzstyle{hedge}=[draw, circle, minimum size=8mm, fill=white];

      \begin{scope}[local bounding box=hypergraph, anchor=center, start chain=going below, node distance=2mm]
        \foreach \i in {1,2,...,6}
        \node[on chain, hedge] (v\i) {$v_\i$}; 
      \end{scope}

      \begin{scope}[draw, anchor=center, shift={(2,-0.1)}, start chain=going below, node distance=8mm]
        \foreach \e/\c in {1/e1_color, 2/e2_color, 3/e3_color, 4/e4_color}
        \node[on chain, draw, circle, fill=\c!50, minimum size=8mm] (e\e) {\textcolor{black}{$e_\e$}};
      \end{scope}

      \draw[e1_color] (v1) -- (e1);
      \draw[e1_color] (v2) -- (e1);
      \draw[e1_color] (v3) -- (e1);

      \draw[e2_color] (v2) -- (e2);
      \draw[e2_color] (v4) -- (e2);

      \draw[e3_color] (v3) -- (e3);
      \draw[e3_color] (v5) -- (e3);
      \draw[e3_color] (v6) -- (e3);

      \draw[e4_color] (v4) -- (e4);
      \draw[e4_color] (v5) -- (e4);
      \draw[e4_color] (v6) -- (e4);

      \draw [decorate, decoration={brace, mirror, amplitude=5pt, raise=4ex}] (v1.north west) -- (v6.south west) node[midway, xshift=-10mm]{\rotatebox{90}{Vertices}};

      \draw [decorate, decoration={brace, amplitude=5pt, raise=4ex}] (e1.north east) -- (e4.south east) node[midway, xshift=10mm]{\rotatebox{90}{Hyperedges}};



    \end{tikzpicture}
    \caption{Any hypergraph can be transformed into an equivalent bipartite graph: The hyperedges and the vertices from the time-varying hypergraph from (a) become the two distinct sets of vertices of a bipartite graph.}
    \label{fig:bipartite_graph}
  \end{subfigure}

  \caption{An example hypergraph (a) and its bipartite-graph equivalent (b).}
  \label{fig:intro}
\end{figure}

Time-varying hypergraphs are a novel concept; therefore, we cannot rely on existing toolings. We implemented the time-hypergraph as an equivalent bipartite graph: The hypergraph vertices and hyperedges become two sets of vertices of the bipartite graph. The vertices of those disjoint sets are connected if a hypergraph edge is part of the hyperedge. \Cref{fig:intro} shows a graphical description of the equivalence of hypergraphs and bipartite graphs.

Since the quality of the experiment and its outcome heavily relies on in-silico experiments, we developed and validated a suitable simulation model for conducting such \emph{in-silico experiments} \cite{Dorner2022}.\footnote{As part of our simulation model, we established the theoretical foundation for a novel generalization of Dijkstra's algorithm for shortest paths in time-varying hypergraphs and its first  implemention. To validate our Python implementation, we developed a extensive test suite in close collaboration with students from a software testing course at Blekinge Institute of Technology, Sweden. We reported our experiences in dedicated study \cite{Dorner2024nofreelunch}.}

We use a modified Dijkstra's algorithm to find the minimal journeys for each vertex (participant) in the time-varying hypergraph. Dijkstra's algorithm is asymptotically the fastest known single-source shortest-path algorithm for arbitrary directed graphs with unbounded non-negative weights. In contrast to its original form, our implementation finds both the shortest (a topological distance) and fastest (a temporal distance) journeys in time-varying hypergraphs.\footnote{For future applications, our implementation of Dijkstra's algorithm can also find any foremost journey.} Since Dijkstra's algorithm can be seen as a generalization of a breadth-first search for unweighted graphs, we can identify not only the minimal paths but also the horizon of each participant in the communication network in one computation.

The algorithm is integrated into our computer model and implemented in Python. For more implementation details and performance considerations, we refer the reader to our replication package, including its documentation \cite{Dorner2023software,Dorner2023data}. Since both time-varying hypergraphs as a data model and the generalized Dijkstra's algorithm for time-varying hypergraphs are novel contributions, we developed an extensive test suite to ensure that the computational model accurately reflects the underlying mathematical framework and that our Dijkstra implementation produces correct results. This effort was carried out in close collaboration with students from a software testing course at BTH. We documented our experiences in a dedicated study \cite{Dorner2024nofreelunch}. Additionally, the model parameterization and simulation code \cite{Dorner2023software} as well as the intermediate and final results \cite{Dorner2023data} are publicly available under an open-source license.

\subsection{Model Parametrization}
\label{sec:parametrization}

Instead of a theoretical or probabilistic parametrization, we parametrize our simulation model with empirical code review systems from three open-code review systems: Android, Visual Studio Code, and React.

In the following, we describe our sampling strategy for selecting suitable open-source code review systems and the code review data extraction process for parametrizing our simulation model.

\subsubsection{Sampling}
\label{sec:sampling}

We use a \emph{maximum variation sampling} to select suitable code review communication networks in open source. A maximum (or maximum heterogeneity) variation sampling is a non-probabilistic, purposive sampling technique that chooses a sample to maximize the range of perspectives investigated in the study in order to identify important shared patterns that cut across cases and derive their significance from having emerged out of heterogeneity \cite{Teddlie2009}.

To ensure comparability with the replicated study, we sampled for the following two perspectives:

\begin{itemize}
  \item \textit{Code review system size}---To avoid a bias introduced by network effects, we required communication networks emerging from different sizes of code review. The size of a communication network can be measured in terms of the number (hyper)edges (corresponding to the number of code reviews) or vertices (corresponding to the number of participants). In our sample, we use a small (React), mid-sized (Visual Studio Code), and large (Android) code review system (see \Cref{tab:sample}). The size classification in small, mid-sized, and large code review systems is arguably arbitrary and relative to the code review systems in our sample rather than following a general norm that, to the best of our knowledge, does not exist.
  \item \textit{Code review tool}---In particular, since \citeauthor{Baum20161} suggested code review tool in use as a main factor shaping code review in industry \cite{Baum20161}, we aim to minimize the code review tool bias for the results and require our sample to contain a diverse set of code review tools. Our sample contains three different code review tools: BitBucket, GitHub, and CodeFlow.
\end{itemize}

In alignment with the qualitative prior work, we explicitly excluded the different manifestations in code review practices as a sampling dimension.

We restrict the population of open-source projects to active and industrial-relevant open-source projects, which are, in our definition, initiated and predominantly led by companies under ongoing development as of this study (at the time of writing this manuscript in 2025). This restriction originates mainly in the heterogeneity of open source: The definition of open-source software refers only to its licensing and does not imply a particular software development method. Thus, there is no single, unified open-source software development and quality assurance process, which makes a comparison of open-source projects inherently difficult. By restricting our population to company-led open-source projects, we aim to reduce this heterogeneity in open-source and make our sample more homogenous.

From this population, we drew a sample of three open-source projects: Android, Visual Studio Code, and React. Those projects are all driven by large software companies, i.e., Google, Microsoft, and Facebook.\footnote{Our replication package includes all necessary software to collect and analyze data from code review systems beyond our sample, ensuring straightforward and efficient replication outside the scope of this study.} \Cref{tab:sample} provides an overview of our sample of code review systems and the dimension of representativeness. We describe the cases in our sample in more detail in the following subsections.

\begin{table}
  \centering
  \caption{Our sample of open-source code review systems with respect to the two dimensions of representativeness during the timeframe under investigation: code review system size and tooling. For completeness and ease of reference, we included the sample of closed-source code review systems from the replicated study~\cite{Dorner2024upperbound}.}
  \label{tab:sample}
  \begin{tabularx}{\textwidth}{@{}Xllll@{}}
    \toprule
    Code review system & Size & Code reviews & Participants & Tooling \\
    \midrule
    \textbf{Open source} \\
    Android & large & \num{10279} & \num{1793} & Gerrit \\
    Visual Studio Code & mid-sized & \num{802} & \num{162} & GitHub \\
    React & small & \num{229} & \num{64} & GitHub \\
    \textbf{Closed source} (via replicated study)  \\
    Microsoft & large & \num{309740} & \num{37103} & CodeFlow \\
    Spotify & mid-sized & \num{22504} & \num{1730} & GitHub \\
    Trivago & small & \num{2442} & \num{364} & BitBucket \\
    \bottomrule
  \end{tabularx}
\end{table}

\paragraph{Android Open Source Project}

The Android Open Source Project (AOSP) is an initiative led by Google to develop and maintain open-source software for mobile devices. It provides the source code for building Android firmware and apps, including the operating system framework, Linux kernel, middleware, and essential system applications. The AOSP serves two purposes: First, as an open-source project, the AOSP offers a platform that is open and accessible to developers, allowing them to customize and extend Android for various devices and use cases. This openness has fostered a vibrant ecosystem of custom ROMs, modifications, and alternative distributions of Android. Second, the AOSP is the reference implementation of \emph{Android}, providing a standard for device manufacturers and developers to follow when creating Android-based products and applications. While Google's own Android releases (such as those on Pixel devices) often include proprietary Google services and applications, AOSP itself is fully open source and can be used as the basis for building Android variants without Google's involvement.
The project uses \emph{Gerrit}\footnote{\url{https://www.gerritcodereview.com}} as code review tool, which mirrored Google's internal code review tool at the time \emph{Mondrian}. The code review within Android Open Source Project is intensively studied \cite{Mukadam2013, Thongtanunam2017, Ebert2017, Thongtanunam2015, Hamasaki2013, Ebert2018, Yang2016, Asri2019, Pascarella2018, Zanjani2016, Thongtanunam2022, Ouni2016, Bavota2015, Rigby2013} which makes it probably the well-known code review system in software engineering research.
Throughout this paper, we will refer to the Android Open Source Project simply as \emph{Android}.

\paragraph{Visual Studio Code}

Visual Studio Code, often abbreviated as VS Code, is a free, open-source code editor initiated and orginally developed by Microsoft. It's widely used by developers for various programming languages and platforms. It's known for its lightweight and fast performance, making it a popular choice among developers for writing code, debugging, and managing projects across different platforms. Visual Studio Code uses GitHub through its pull-request functionality as code review tool.

\paragraph{React}

React is a popular JavaScript library for building user interfaces, particularly for web applications. The open-source project was initiated and is led by Facebook. React is known for its declarative and component-based approach to building UIs. It allows developers to create reusable UI components that manage their own state, making it easier to build complex UIs with minimal code duplication. %
Like Visual Studio Code, React uses GitHub through its pull-request functionality as code review tool.

\subsubsection{Data Collection}
\label{sec:data_collection}

We extract all human interactions with the code review discussions within four consecutive calendar weeks from the single, central code review tools in each open-source software development context.

We define a code review interaction as any non-trivial contribution to the code review discussion: creating, editing, approving or closing, and commenting on a code review. For this study, we do not consider other (tool-specific) types of discussion contributions (for example, emojis or likes) a substantial contribution to a code review.

The beginning and end of those four-week timeframes differ and are arbitrary, but share the common attributes: All timeframes

\begin{itemize}
  \item start on a Monday and end on a Sunday,
  \item have no significant discontinuities by public holidays such as Christmas or Midsommar,
  \item are pre-pandemic to avoid introduced noise from changes in remote work policies, pandemic-related restrictions, or interferences in the software development.
\end{itemize}

\Cref{tab:timeframes} lists the timeframes (each four weeks) and when the data was collected.

\begin{table}
  \centering
  \caption{Timeframes and the data collection timeframe among our cases. }
  \label{tab:timeframes}
  \begin{tabular}{lll}
    \toprule
    & Timeframe (four weeks) & Collected during \\
    \midrule
    Android & 2024-03-04 to 2024-03-31 & April 2025\\
    Visual Studio Code & 2024-03-04 to 2024-03-31 & April 2025\\
    React & 2024-03-04 to 2024-03-31 & April 2025\\
    \bottomrule
  \end{tabular}
\end{table}

We excluded all non-human code-review participants and interactions (i.e., bots or automated tasks contributing to the code-review discussions). To protect the privacy of all individuals, we strictly anonymized all participants and removed all identifiable personal information.

We made all data and results, as well as the extraction pipeline for Gerrit and GitHub, publicly available.

\section{Results}
\label{sec:results}

This section presents four measurements from our simulation, described in \Cref{sec:experimental_design}, structured around our two research questions: how widely (RQ~1) and how quickly (RQ~2) information can spread in code review under best-case assumptions.

To address RQ~1, we present the measurements on the absolute (\Cref{sec:normalized_information_diffusion_range}) and normalized information diffusion range (\Cref{sec:absolute_information_diffusion_range}), capturing how far information can reach across participants. For RQ~2, we examine the topological (\Cref{sec:topological_distances}) and temporal distances between participants (\Cref{sec:temporal_distances}), reflecting the speed of potential information diffusion in terms of structure and time.

To enable clearer and more direct comparisons, we present the results from the open-source code review systems alongside findings from a replicated study that examines the upper bound of information diffusion in closed-source code review systems at large (Microsoft), mid-sized (Spotify), and small (Trivago) companies. The colors representing the companies is consistent throughout this paper. Values in figures and tables are normalized to the $[0, 1]$ range, while percentages are used in the text for readability.

\subsection{How wide can information spread within code review (RQ~1)?}

As outlined in \Cref{sec:conceptual_model}, we address RQ~1 by measuring how many participants each individual can reach in the code review communication network. Mathematically, this is the cardinality of a participant’s horizon, which we refer to as the \emph{information diffusion range}. We report both the normalized and absolute diffusion range.

\subsubsection{Normalized Information Diffusion Range}
\label{sec:normalized_information_diffusion_range}
To make the different code review system sizes comparable, we normalize the information diffusion range to the number of code review participants in a code review system. Mathematically, we define the normalized information diffusion range for all code review participants $u \in V$ by

\[
  \frac{\lvert \{v \in V \colon u \leadsto v \} \rvert}{|V|}.
\]

\begin{figure}
  \centering
  \begin{tikzpicture}
    \begin{axis} [ %
        ecdf axis,
        xmin=0.0,
        xmax=1.0,
        xlabel={Normalized information diffusion range},
        ylabel={Fraction of normalized information diffusion ranges $\leq x$},
        legend pos=north west,
      ]
      \addplot[android_color, name path=androidcov] table [x=index, y=Android, col sep=comma] {data/csv/ranges.csv} coordinate (maxandroid);
      \addlegendentry{Android}
      \addplot[vscode_color, name path=vscodecov] table [x=index, y=Visual Studio Code, col sep=comma] {data/csv/ranges.csv} coordinate (maxvscode);
      \addlegendentry{Visual Studio Code}
      \addplot[react_color, name path=reactcov] table [x=index, y=React, col sep=comma] {data/csv/ranges.csv} coordinate (maxreact);
      \addlegendentry{React}

      \addplot[microsoft_color, name path=microsoftcov] table [x=index, y=Microsoft, col sep=comma] {data/csv/ranges.csv} coordinate (maxmicrosoft);
      \addlegendentry{Microsoft}
      \addplot[spotify_color, name path=spotifycov] table [x=index, y=Spotify, col sep=comma] {data/csv/ranges.csv} coordinate (maxspotify);
      \addlegendentry{Spotify}
      \addplot[trivago_color, name path=trivagocov] table [x=index, y=Trivago, col sep=comma] {data/csv/ranges.csv} coordinate (maxtrivago);
      \addlegendentry{Trivago}

    \end{axis}
  \end{tikzpicture}
  \caption{Empirical cumulative distribution of the normalized information diffusion ranges per code review system after four weeks.}
  \label{fig:normalized_ranges}
\end{figure}
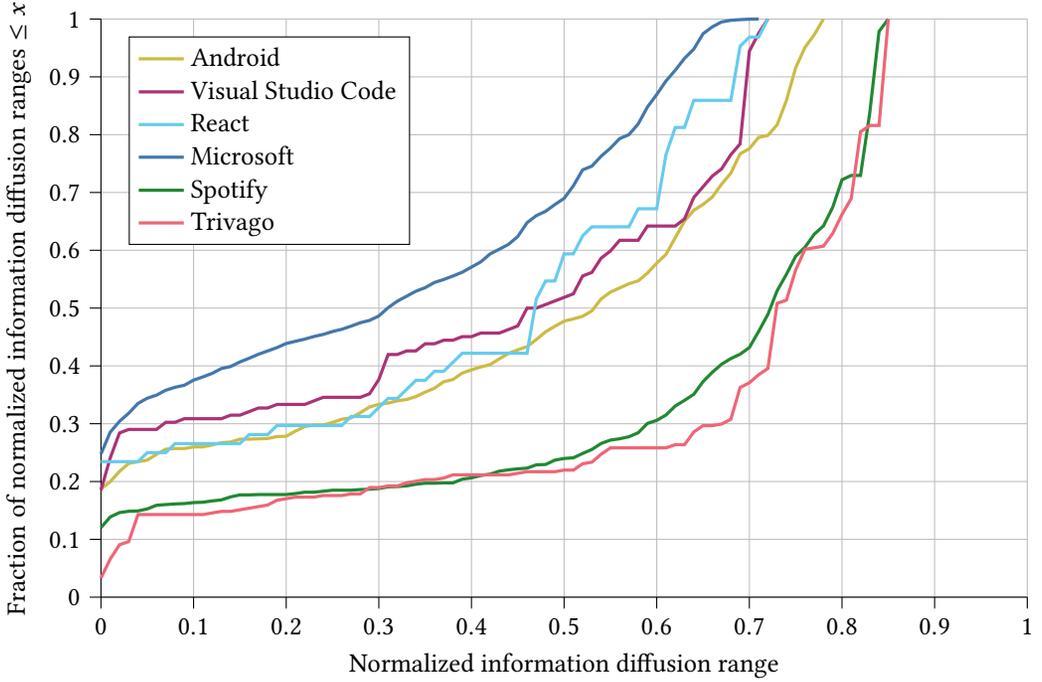

\Cref{fig:normalized_ranges} plots the empirical cumulative distribution functions (ECDF) visualizing the distributions of the normalized information diffusion range per participant after four weeks each resulting from our simulation.

We identified the upper bound of normalized information diffusion range within closed-source code review systems at both Trivago and Spotify. Interestingly, the two platforms exhibit nearly identical distributions in terms of how widely participants can spread information relative to the size of their respective networks. Specifically, at Trivago, a code review participant can reach up to \SI{85}{\percent} of their network. Furthermore, \SI{30}{\percent} of participants are able to reach between \SI{81}{\percent} and \SI{85}{\percent} of all participants, while the median participant can reach between \SI{72}{\percent} and \SI{85}{\percent} of all participants within the network.

No open-source code review system exceeded the upper bound found in the closed-source code review systems---lending strong support to the findings of the replicated study. We observed the highest normalized information diffusion range in open-source systems at Android: A code review participants in the code review systems of Android can reach \qty{78}{\percent} of its network at maximum. For a detailed comparison, we refer to \Cref{tab:normalized_ranges}, which presents the upper quantile ranges of information diffusion---specifically, the \numrange{0.3}{1.0}, \numrange{0.5}{1.0}, \numrange{0.7}{1.0}, and \numrange{0.9}{1.0} ranges---in each code review system.

\begin{table}
  \centering
  \caption{Quantile ranges of the normalized information diffusion ranges per participant for each code review system.}
  \label{tab:normalized_ranges}
  \begin{tabular}{lllll}
\toprule
 & \multicolumn{4}{l}{Quantile range} \\
 & \numrange{0.3}{1.0} & \numrange{0.5}{1.0} & \numrange{0.7}{1.0} & \numrange{0.9}{1.0} \\
\midrule
React & \numrange{0.19}{0.72} & \numrange{0.39}{0.72} & \numrange{0.58}{0.72} & \numrange{0.64}{0.72} \\
Visual Studio Code & \numrange{0.03}{0.72} & \numrange{0.45}{0.72} & \numrange{0.64}{0.72} & \numrange{0.69}{0.72} \\
Android & \numrange{0.23}{0.78} & \numrange{0.53}{0.78} & \numrange{0.66}{0.78} & \numrange{0.74}{0.78} \\
Microsoft & \numrange{0.01}{0.71} & \numrange{0.30}{0.71} & \numrange{0.50}{0.71} & \numrange{0.61}{0.71} \\
Spotify & \numrange{0.58}{0.85} & \numrange{0.72}{0.85} & \numrange{0.79}{0.85} & \numrange{0.83}{0.85} \\
Trivago & \numrange{0.67}{0.85} & \numrange{0.72}{0.85} & \numrange{0.81}{0.85} & \numrange{0.83}{0.85} \\
\bottomrule
\end{tabular}

\end{table}

Although \Cref{fig:normalized_ranges} reveals a comparable pattern in the normalized information diffusion range---reflecting how widely information can spread during code review---in the three open-source code review systems, we could not find a clear explanation for them. Neither the network size (measured by the number of code reviews or participants) nor the specific code review tool accounts for these patterns. For example, despite the significant difference in size (\num{37104} code review participants at Microsoft and \num{162} code review participants at Visual Studio Code), the normalized information diffusion range in Microsoft’s closed-source code review systems and the open-source system for Visual Studio Code shows a comparable distribution of reachable participants relative to system size. With a median of \qty{45}{\percent} and \qty{72}{\percent} at Visual Studio Code and \qty{30}{\percent} and \qty{71}{\percent} of reachable participants, both code review systems define the smallest diffusion range in our sample.

\subsubsection{Absolute Information Diffusion Range}
\label{sec:absolute_information_diffusion_range}

If we consider the absolute information diffusion range for each code instead, which is defined for all code review participants $u \in V$ by

\[\lvert \{v \in V \colon u \leadsto v \} \rvert, \]

code review participants at Microsoft's code review system can reach the most participants. Although Microsoft's code review system, along with Visual Studio Code's, exhibits the smallest normalized information diffusion range, it sets the upper bound for the absolute information diffusion range in our sample. In detail, the code review system at Microsoft can spread information up to \num{26216} participants (\qty{71}{\percent} of the total network size), half of the code review participants can reach \num{11645} or more other participants.\Cref{tab:absolute_ranges} lists the ranges of the absolute information diffusion range we found for the quantile ranges \numlist{0.7;0.5;0.3;0.1}

\begin{table}
  \centering
  \caption{Quantile ranges of the absolute information diffusion ranges per participant for each code review system.}
  \label{tab:absolute_ranges}
  \begin{tabular}{lllll}
\toprule
 & \multicolumn{4}{l}{Quantile range} \\
 & \numrange{0.3}{1.0} & \numrange{0.5}{1.0} & \numrange{0.7}{1.0} & \numrange{0.9}{1.0} \\
\midrule
React & \numrange{12}{46} & \numrange{25}{46} & \numrange{37}{46} & \numrange{41}{46} \\
Visual Studio Code & \numrange{5}{116} & \numrange{74}{116} & \numrange{104}{116} & \numrange{113}{116} \\
Android & \numrange{448}{1407} & \numrange{966}{1407} & \numrange{1197}{1407} & \numrange{1344}{1407} \\
Microsoft & \numrange{808}{26216} & \numrange{11645}{26216} & \numrange{18887}{26216} & \numrange{22983}{26216} \\
Spotify & \numrange{1026}{1472} & \numrange{1260}{1472} & \numrange{1386}{1472} & \numrange{1447}{1472} \\
Trivago & \numrange{245}{310} & \numrange{266}{310} & \numrange{296}{310} & \numrange{309}{310} \\
\bottomrule
\end{tabular}

\end{table}

Among open source projects, the upper bound of the absolute information diffusion range is substantially smaller. For example, half of the participants in Android's code review system are exposed to information from between \num{966} and \num{1407} other participants. Even considering the exceptionally large size at Microsoft, Android and Spotify have a comparable number of code review participants (\num{1793} at Android and \num{1730} at Spotify), the absolute information diffusion range is notably more constrained in Android.

\subsection{How quickly can information spread within code review (RQ~2)?}

As outlined in \Cref{sec:conceptual_model}, we address RQ~2 by analyzing the distances between code review participants. In time-varying hypergraphs, distance is characterized in two distinct ways: topological distance, defined as the minimal number of hops across all time-respecting paths (or journeys), and temporal distance, defined as the minimal duration of those time-respecting paths (or journeys).

Accordingly, our answers to RQ~2 are structured around these two complementary notions of distance.

\subsubsection{Topological Distances in Code Review}
\label{sec:topological_distances}

\begin{figure}
  \centering
  \begin{tikzpicture}
    \begin{semilogxaxis}[
        ecdf axis,
        log basis x=2,
        xmin=1,
        xmax=39,
        legend pos=south east,
        xlabel={Topological distance},
        ylabel={Fraction of distances $\leq x$},
        xtick={1, 2, 3, 4, 5, 6, 7, 14, 20, 39},
        xticklabels={1, 2, 3, 4, 5, 6, 7, 14, 20, 39},
        extra x ticks={1, 2, 3, 4, 5, 6, 7, 8, 9, 10, 11, 12, 13, 14, 15, 16, 17, 18, 19, 20, 21, 22, 23, 24, 25, 26, 27, 28, 29, 30, 31, 32, 33, 34, 35, 36, 37, 38, 39},
        extra x tick labels={},
        extra x tick style={
          major tick length=0pt
        }
      ]

      \addplot[android_color, name path=androidshortest] table [x=index, y=Android, col sep=comma] {data/csv/topological_distances.csv} coordinate (maxandroid);
      \addlegendentry{Android}
      \addplot[vscode_color, name path=vscodeshortest] table [x=index, y=Visual Studio Code, col sep=comma] {data/csv/topological_distances.csv} coordinate (maxvscode);
      \addlegendentry{Visual Studio Code}
      \addplot[react_color, name path=reactshortest] table [x=index, y=React, col sep=comma] {data/csv/topological_distances.csv} coordinate (maxreact);
      \addlegendentry{React}

      \addplot[microsoft_color, name path=microsoftshortest] table [x=index, y=Microsoft, col sep=comma] {data/csv/topological_distances.csv} coordinate (maxmicrosoft);
      \addlegendentry{Microsoft}
      \addplot[spotify_color, name path=spotifyshortest] table [x=index, y=Spotify, col sep=comma] {data/csv/topological_distances.csv} coordinate (maxspotify);
      \addlegendentry{Spotify}
      \addplot[trivago_color, name path=trivagoshortest] table [x=index, y=Trivago, col sep=comma] {data/csv/topological_distances.csv} coordinate (maxtrivago);
      \addlegendentry{Trivago}

    \end{semilogxaxis}
  \end{tikzpicture}
  \caption{Empirical cumulative distribution of the topological distances between participants per code review system after four weeks. The topological distance is the minimal number of code reviews (hops) required to spread information from one code review participant to another.}
  \label{fig:topological_distances}
\end{figure}
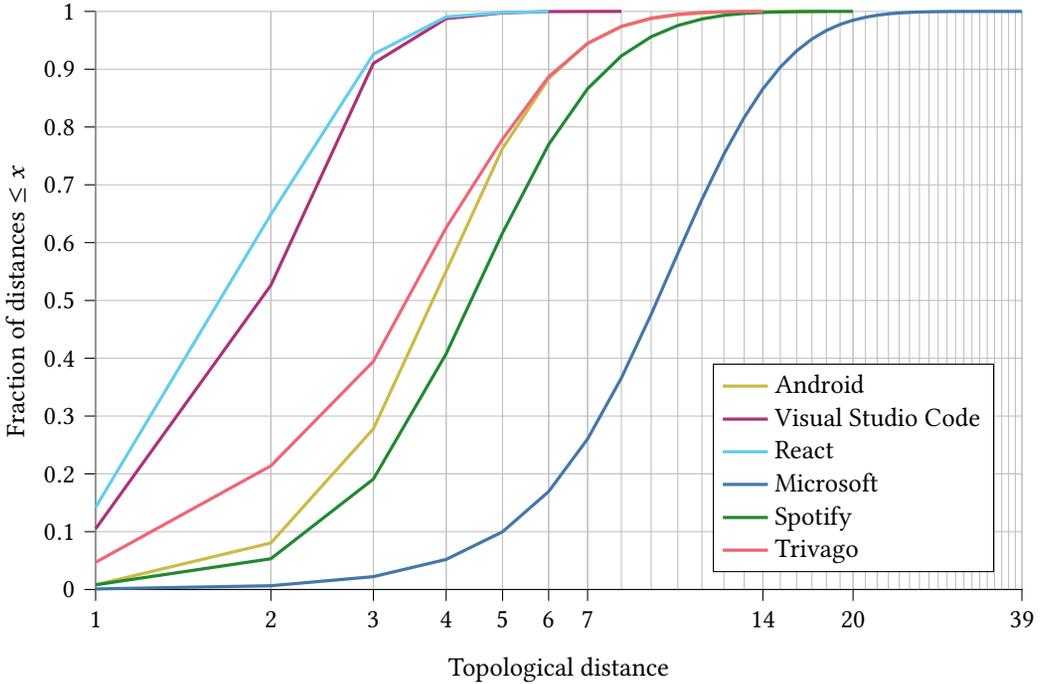

\Cref{fig:topological_distances} illustrates the empirical cumulative distribution of topological distances among code review participants in the analyzed code review systems.

In comparison to our prior findings, the open-source code review systems of React and Visual Studio Code extend the upper bound of how quickly information can spread through code review. Specifically, the average (median) distance between two arbitrary participants is one for React and Visual Studio Code, for Android it is three. In contrast, the average (median) distances at Trivago, Spotify, and Microsoft are three, four, and nine\footnote{In the replicated study, we inadvertently reported Microsoft's topological distances with an off-by-one error. Although we consider this a minor issue that does not impact the validity of the previous findings, we believe it is important to explicitly acknowledge the error. We have verified the results and corrected it in this paper to ensure completeness and accuracy.} hops, respectively. The maximum observed distances are 14 hops for Trivago, 20 hops for Spotify, and 39 hops for Microsoft\footnotemark[\value{footnote}]. Open source code review systems again are significant shorter and have a maximum observed distance of 18 (for Android), 8 (for Visual Studio Code), and 6 (for React).

\subsubsection{Temporal Distances in Code Review}
\label{sec:temporal_distances}

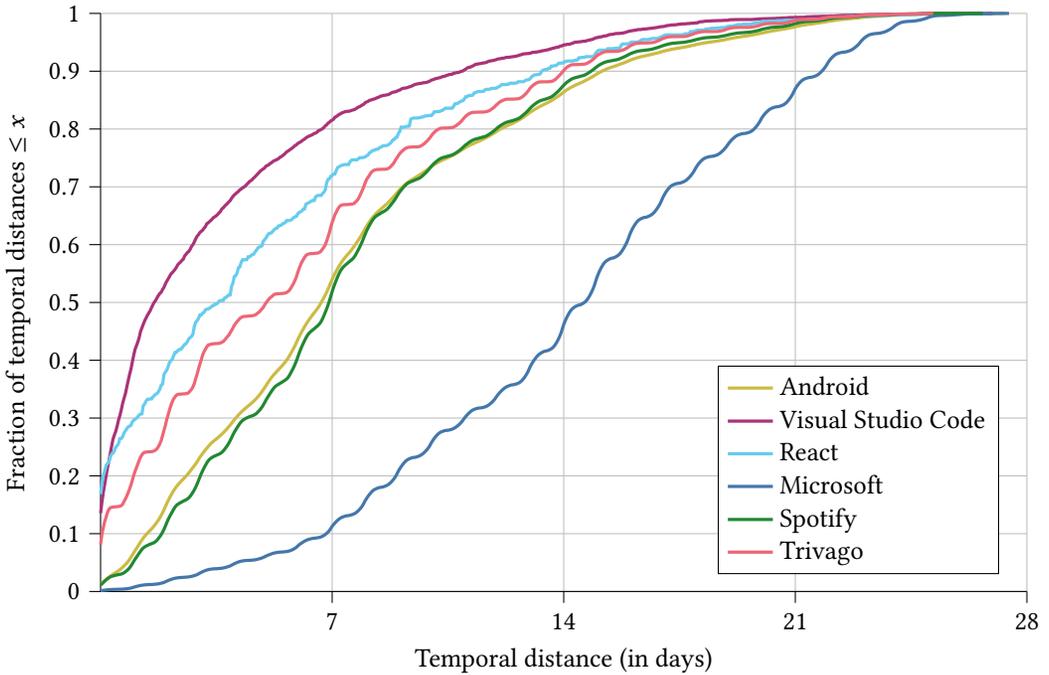
\begin{figure}
  \centering
  \begin{tikzpicture}
    \begin{axis}[
        ecdf axis,
        xmax=672,
        xtick={168, 336, 504, 672},
        xticklabels={7, 14, 21, 28},
        legend pos=south east,
        xlabel={Temporal distance (in days)},
        ylabel={Fraction of temporal distances $\leq x$},
      ]

      \addplot[android_color, name path=android] table [x=index, y=Android, col sep=comma] {data/csv/temporal_distances.csv};
      \addlegendentry{Android}
      \addplot[vscode_color, name path global=vscode] table [x=index, y=Visual Studio Code, col sep=comma] {data/csv/temporal_distances.csv};
      \addlegendentry{Visual Studio Code}
      \addplot[react_color, name path=react] table [x=index, y=React, col sep=comma] {data/csv/temporal_distances.csv};
      \addlegendentry{React}
      \addplot[microsoft_color, name path=microsoftcov] table [x=index, y=Microsoft, col sep=comma] {data/csv/temporal_distances.csv};
      \addlegendentry{Microsoft}
      \addplot[spotify_color, name path global=spotifycov] table [x=index, y=Spotify, col sep=comma] {data/csv/temporal_distances.csv};
      \addlegendentry{Spotify}
      \addplot[trivago_color, name path=trivagocov] table [x=index, y=Trivago, col sep=comma] {data/csv/temporal_distances.csv};
      \addlegendentry{Trivago}
    \end{axis}
  \end{tikzpicture}
  \caption{The cumulative distribution of minimal temporal distances between participants in code review systems. The temporal distance is the minimal duration required to spread information from one participant to another.}
  \label{fig:temporal_distances}
\end{figure}

The other type of distance in time-varying hypergraphs is temporal distance. The fastest time-varying path refers to the path between two code review participants that minimizes the (relative) temporal distance; i.e., the shortest timespan required for information to spread from one participant to another. Given our observation window, the temporal distance in our measurements cannot exceed four weeks. \Cref{fig:temporal_distances} shows the cumulative distribution of the relative temporal distances between the code review participants in our sample.

Again, an open-source code review pushes the boundaries of how quickly information can spread in code review: On average, information can spread through the code review system of Visual Studio Code in less than a day and at React less than three days. The average (median) temporal distance between two code review participants at Android, Trivago, and Spotify is less than seven days, while a code review participant a Microsoft takes more than 14 days, which is still in the observation window of four weeks.

From \Cref{fig:temporal_distances}, we observe that the temporal distances in closed-source code review systems are tightly constrained to a typical workday schedule. All closed-source code review systems exhibit distinct step patterns in their ECDFs, reflecting a pronounced day-night rhythm. In contrast, the ECDFs for all open-source code review systems are notably smoother, suggesting a more continuous and globally distributed review activity with less coupling to traditional workday boundaries.

\section{Discussion}
\label{sec:discussion}

As outlined in the experimental design in \Cref{sec:experimental_design}, we now turn to a discussion of the results as evidence testing the theory that code review would function effectively as a communication network. In other words, do our four measurements indicate that code review fails to enable information diffusion at scale, thereby calling into question its role as an effective communication medium for information exchange among developers?

\begin{figure}

  \begin{tikzpicture}[
      x=5.1cm,
      y=1.25cm,
      every node/.style={
        font=\small,
      },
      measurement/.style={
        draw,
        rounded corners,
        align=center,
        text width=3cm,
        font=\small,
      },
      rq/.style={
        draw,
        rounded corners,
        align=center,
        text width=3cm,
        font=\small,
      },
      h/.style={
        draw,
        rounded corners,
        align=center,
        text width=3.35cm,
        font=\small,
      },
      cat/.style={
        align=center,
        font=\bfseries,
    }]

    \node[cat] at (0,2) {Measurements};
    \node[cat] at (1,2) {Research Questions};
    \node[cat] at (2,2) {Hypotheses};

    \node[measurement] (m1) at (0,0.5) {Normalized information diffusion range};
    \node[measurement] (m2) at (0,-0.5) {Absolute information diffusion range};
    \node[measurement] (m3) at (0,-2) {Topological distance between participants};
    \node[measurement] (m4) at (0,-3) {Temporal distance between participants};

    \node[rq] (rq1) at (1,0) {RQ~1: How widely can information spread within [code review]?};
    \node[rq] (rq2) at (1,-2.5) {RQ~2: How quickly can information spread within [code review]?};

    \node[h] (h1) at (2,0) {$H_1$: Code review spreads information quickly.};
    \node[h] (h2) at (2,-2.5) {$H_2$: Code review spreads information widely.};
    \node[draw, dashed, rounded corners, fit=(h1)(h2), inner sep=5pt, label={[text width=5cm, align=center, font=\itshape]below:Theory of code review as communication network}] {};

    \draw[-latex] (m1) -- node[midway, above] {answers} (rq1);
    \draw[-latex] (m2) -- node[midway, above] {answers} (rq1);
    \draw[-latex] (rq1) -- node [pos=0.5, above] {tests} (h1);

    \draw[-latex] (m3)  -- node[midway, above] {answers} (rq2);
    \draw[-latex] (m4)  -- node[midway, above] {answers} (rq2);
    \draw[-latex] (rq2) -- node [pos=0.5, above] {tests} (h2);

  \end{tikzpicture}
  \caption{Summary of the four diffusion measurements used in the study, their alignment with the research questions RQ~1 and RQ~2, and references to the corresponding hypothesis.}
  \label{fig:evaluation}
\end{figure}
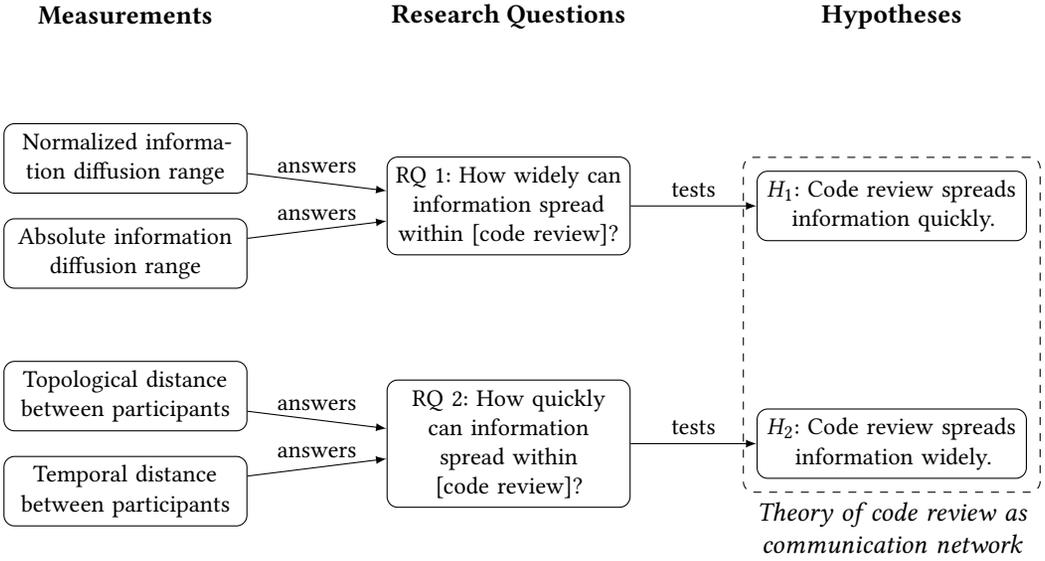

\Cref{fig:evaluation} summarizes and highlights the interdependencies among the four measurements, the two research questions, and the two hypotheses introduced in \Cref{sec:experimental_design}.

\subsection{Code review can spread information widely among its participants (Hypothesis~1)}

Our results provide strong support for the first hypothesis, particularly in closed-source environments. Participants in code review at Trivago or Spotify, for example, can reach up to 85\% of their network under best-case conditions (cf. \Cref{fig:normalized_ranges}), and a large proportion of participants achieve high diffusion ranges. These results reflect dense, tightly integrated communication structures that promote high reachability. In terms of absolute information diffusion range, the closed-source system at Microsoft stands out with participants reaching up to 26,216 others, even though its normalized range is relatively low due to its sheer scale (cf. \Cref{tab:absolute_ranges}).

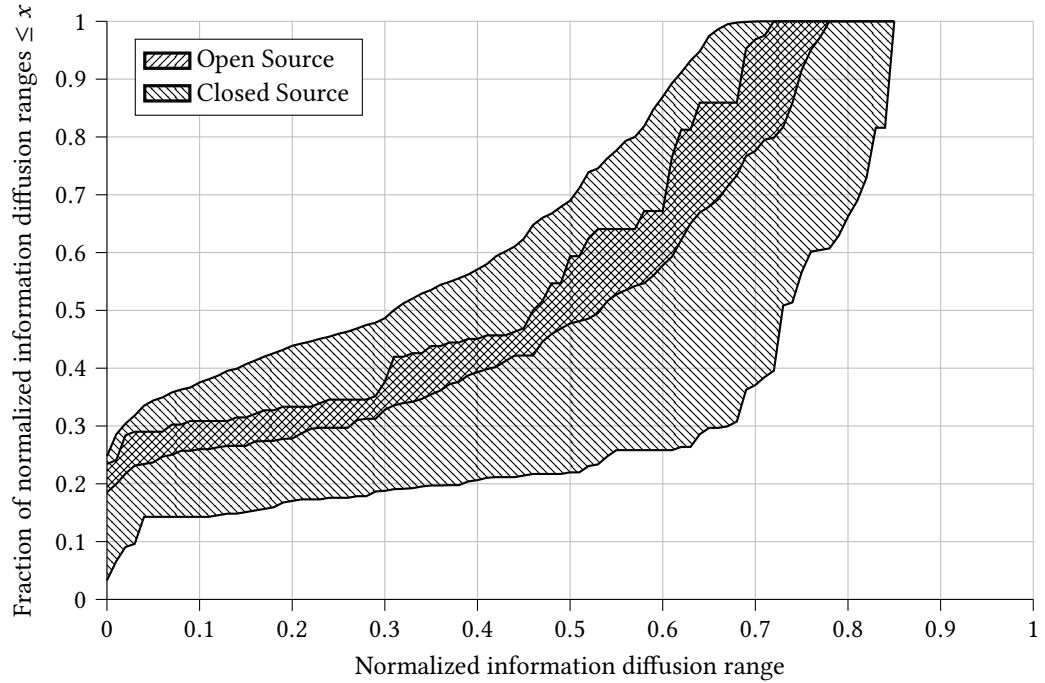
\begin{figure}
  \centering
  \begin{tikzpicture}
    \begin{axis} [ %
        ecdf axis,
        xmin=0.0,
        xmax=1.0,
        xlabel={Normalized information diffusion range},
        ylabel={Fraction of normalized information diffusion ranges $\leq x$},
        legend pos=north west,
        legend entries={,,Open Source,,,Closed Source},
      ]
      \addplot[thick, name path=osmin] table [x=index, y=os_min, col sep=comma] {data/csv/ranges.csv};
      \addplot[thick, name path=osmax] table [x=index, y=os_max, col sep=comma] {data/csv/ranges.csv};
      \addplot[ pattern=north east lines, pattern color=black,] fill between[of=osmin and osmax];

      \addplot[thick, name path=csmin] table [x=index, y=cs_min, col sep=comma] {data/csv/ranges.csv};
      \addplot[thick, name path=csmax] table [x=index, y=cs_max, col sep=comma] {data/csv/ranges.csv};
      \addplot[pattern=north west lines,pattern color=black,] fill between[of=csmin and csmax];

    \end{axis}
  \end{tikzpicture}
  \caption{The aggregated empirical cumulative distributions of the normalized information diffusion ranges in open-source and closed-source code review systems after four weeks.}
  \label{fig:normalized_ranges_diff}
\end{figure}

However, this capability does not translate uniformly across environments. Our simulation revealed that open-source code review systems consistently fall short of the upper bounds set by closed-source systems in terms of how widely information can spread (cf. \Cref{fig:normalized_ranges_diff}). Even popular and highly relevant open-source projects such as React and Android only approach but do not exceed the normalized diffusion range seen in closed-source systems.

Our results also identified three distinct diffusion patterns across systems (cf. \Cref{fig:normalized_ranges}), yet no simple variable (e.g., system size, tooling, or openness) explained these differences. For instance, the median reach in Android is less than 1,500 participants, while Spotify’s median exceeds this, despite similar number of participants (cf. \Cref{tab:absolute_ranges}). These results suggest that structural factors---such as internal communication protocols or enforced participation policies---may facilitate wider diffusion in closed-source environments.

Thus, while $H_1$ is supported overall, the degree of diffusion is strongly modulated by whether the system is open or closed source. Closed-source code review systems consistently enable wider diffusion. We speculate that the underlying organizational dynamics play a critical role in shaping these outcomes. In particular, closed-source systems may support broader information diffusion because they tend to operate in more decentralized review environments, where code review responsibilities are distributed across multiple engineers and teams. In contrast, open-source projects often centralize review activity in a small set of core maintainers, who act as gatekeepers and conduct the majority of reviews. This centralized control, combined with limited social or organizational trust toward transient or unknown contributors, may constrain the pathways through which information can spread.

\subsection{Code review can spread information quickly among its participants (Hypothesis~2)}

The results also support hypothesis $H_2$: information can spread quickly among code review participants, particularly in open-source environments. In terms of topological distance, which measures the number of hops between participants, open-source systems such as React and Visual Studio Code define the shortest communication paths across our sample (cf. \Cref{fig:topological_distances}). In React, the average path length between any two participants is just one hop, indicating high connectivity and short transmission chains.

These structural observations are reinforced by measurements of temporal distance---how long it takes for information to spread along time-respecting paths. Again, React sets the benchmark: on average, information can reach any participant in less than 3.5 days (cf. \Cref{fig:temporal_distances}). In contrast, Microsoft's system---despite its large reach---exhibits the longest average delay (over 14 days), suggesting that scale may come at the cost of immediacy.

Temporal activity patterns further differentiate the systems. Open-source code reviews exhibit smoother ECDFs of temporal distance, with less evidence of the day-night cycles seen in closed-source systems (cf. \Cref{fig:temporal_distances}). This likely reflects the distributed and asynchronous nature of open-source contributions, where global participation enables continuous communication, independent of localized work schedules.

Together, the topological and temporal distance results demonstrate that, under idealized conditions, information can not only spread quickly through the code review network but may do so more efficiently in decentralized and open environments. At the same time, our findings also reinforce our previous speculation: that centralized review structures---common in open-source projects---may contribute to shorter topological paths by concentrating review activity through a small number of highly active core maintainers. These maintainers effectively act as communication hubs, reducing the number of hops needed for information to flow between otherwise unconnected participants. While this centralized structure may constrain the overall diffusion range, it can facilitate rapid spread within a smaller core network. In contrast, closed-source systems often distribute review responsibilities more broadly across teams, which supports wider diffusion but can result in longer average paths. Future work should explore this trade-off more systematically, by empirically quantifying the degree of review centralization and its relationship to both diffusion speed and reach across diverse organizational and community contexts.

\section{Implications}

In this section, we explore the implications and broader value of conceptualizing code review as a communication network that is grounded in a stronger theoretical foundation. Our aim is to show how a well-supported theory can function not only as an analytical tool, but also as a vehicle for advancing our understanding of collaborative software engineering. While many implications may follow from this perspective, we highlight three that we believe are particularly significant---for research and for practice---each demonstrating how this theory can inform future directions in software engineering.

\subsection{Code Review as Proxy for Collaborative Software Engineering}

If code review functions as a communication network among developers it can serve as a measurable and reliable proxy for the---potentially very fine-grained---collaboration. While insights into collaborative software engineering have wide-ranging applications in research and practice, the collaboration within a multinational enterprise across national borders has an often overlooked legal implication: the profits from those cross-border collaborations within a multinational enterprise become taxable \cite{Dorner2024tax}. Given that tax compliance presents a significant financial risk to multinational enterprises---as highlighted by the ongoing case involving Microsoft \cite{Treidler2024}---it is imperative that multinational enterprises rely on robust and reliable measurements, grounded in a solid theoretical framework, to ensure accurate reporting to tax authorities and to mitigate the risks associated with non-compliance. In our work \cite{Dorner2024tax}, we suggest cross-border code reviews---instances where participants are employed by legal entities in different countries---at large multinational as proxy for collaborative software engineering across international borders and lay the foundation for a new line of research on tax compliance in collaborative software engineering \cite{Dorner2025esem}.

\subsection{Differences Between Open-Source and Closed-Source Code Review}

Our findings reveal notable differences in how open-source and closed-source code review systems function as communication networks. These differences are not merely a matter of scale, but reflect deeper variations in organizational structure, governance, and review practices. Open-source projects often exhibit more centralized review activity around a small group of maintainers \cite{Bosu2017, Rigby2012, Alami2019}, while closed-source systems distribute review responsibilities more broadly across formal teams with shared accountability \cite{Sadowski2018, Smite2023}. These distinct patterns could explain the substantial differences in the capabilities of code review as communication network in open-source and closed source software development.

As a result, care must be taken when interpreting empirical findings across these contexts. Many prior studies rely on open-source data---often due to its accessibility and transparency---but our results suggest that insights gained from open-source systems may not generalize to closed-source environments without further considerations. Researchers and practitioners should be cautious in transferring assumptions or models derived from one setting to another without considering structural differences.

But rather than viewing these differences as limitations, we see them as opportunities to develop a more nuanced understanding of code review as a socio-technical process. The apparent strengths of both open-source (speed) and closed-source (reach) environments suggest the potential value of hybrid models like inner source \cite{Capraro2016}.

\subsection{Impact of AI on the Communicative Role of Code Review}

A more forward-looking implication of our work concerns the evolving role of code review in light of rapid advances in AI-assisted development. As automated tools increasingly contribute to code generation, validation, and even code review, the practice of code review is likely to undergo substantial transformation.

While deterministic bots have long played a role in code review---for example, in reporting testing results or code quality characteristics---the emergence of non-deterministic, generative AI systems introduces a qualitatively different shift. Unlike rule-based tools, generative AI can produce context-sensitive and novel responses, potentially functioning as autonomous participants in code review, both as reviewer or code author. As these systems might transition from assistants to co-reviewers or even sole reviewers, they may begin to reshape the communicative and collaborative function of code review itself.

We anticipate three key types of erosion that may ultimately signal the end of code review as we know it today. First, \emph{erosion of understanding}: AI-generated code and reviews may prioritize machine comprehension over human readability, reducing developers' ability to maintain and reason about the codebase. Second, \emph{erosion of accountability}: as AI systems cannot be held responsible for the changes they influence, gaps in ownership and regulatory compliance may emerge. Third, \emph{erosion of trust}: the blurring of authorship between humans and AI risks undermining confidence in the authenticity and reliability of the review process.

Together, these concerns suggest that deeper integration of generative AI into code review may not simply improve efficiency but fundamentally alter---or even dismantle---the collaborative, human-centered foundation of code review. Without careful oversight, we may need to radically rethink its role in future software development ecosystems, both in research and practice.

\section{Threats to Validity}
\label{sec:threats_to_validity}

As with any empirical study, our findings are subject to limitations that may influence the strength, scope, or interpretation of our results. To support transparency and enable critical assessment, we reflect on the potential threats to the validity of our study, following established frameworks in empirical software engineering. Specifically, we discuss threats to internal validity (whether our results are caused by the study design rather than confounding factors), construct validity (whether our measurements accurately reflect the concept of information diffusion), external validity (the extent to which our findings generalize to other settings), and conclusion validity (the reliability and potential bias in our inferences and interpretations). Where possible, we outline strategies to mitigate these threats and suggest directions for future work to strengthen the robustness of the research.

\subsection{Internal Validity}

Our study simulates information diffusion under best-case conditions. However, several factors may threaten internal validity---that is, whether the observed outcomes can be confidently attributed to our simulation logic rather than to uncontrolled or confounding variables.

First, the boundary between open-source and closed-source development introduces a structural asymmetry that may affect internal consistency. In closed-source environments, the development context is clearly bounded: contributors are typically employees, and communication occurs within the confines of an organization. In contrast, open-source projects often lack a well-defined membership boundary. Developers may participate across multiple projects, both professionally and voluntarily, and information frequently flows between interrelated codebases. To ensure a consistent analytical scope, we rely on the boundaries defined by the code review tool and its configuration. While technically consistent, this boundary may be artificial---particularly in open-source settings where review activity and knowledge sharing extend across repositories. As a result, our simulation may underestimate the actual diffusion potential in open ecosystems. Future work could address this limitation by modeling open-source environments as interlinked review networks rather than isolated systems.

Second, a structural limitation arises from the intrinsic differences in scale between open-source and closed-source code review systems. Open-source projects, by nature, rarely reach the scale of corporate systems. Even the largest open-source systems in our sample (e.g., Android or Visual Studio Code) involve only a few hundred active participants, whereas Microsoft’s internal review network includes over 37,000 contributors (cf. \Cref{tab:sample}). These differences are not incidental but reflect fundamental constraints of each ecosystem. Consequently, comparing diffusion capacity across systems of such disparate scale risks conflating size with capability. Closed-source systems are simply able to support broader diffusion because they operate at a fundamentally different scale. Even with normalized metrics, the raw potential for information to reach more participants in large systems may amplify or mask structural effects that are not present in smaller, open networks. Thus, any comparison of diffusion between open-source and closed-source systems must be interpreted with caution: what appears to be a performance difference may instead reflect the structural ceiling each system type can naturally support.

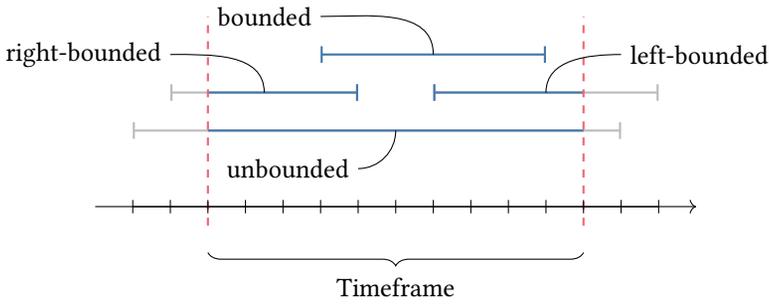
\begin{figure}
  \centering
  \tikzstyle{cr}=[draw, |-|, thick, gray]
  \begin{tikzpicture}[x=1cm, y=0.5cm]

    \draw[cr, blue] (4.5,3) -- (7.5,3) node[midway] (bounded) {};

    \draw[cr] (2.5, 2) -- (5, 2) node[midway] (right_bounded) {};
    \draw[draw, thick, -|, blue] (3, 2) -- (5, 2);

    \draw[cr] (6,2) -- (9,2) node[midway] (left_bounded) {};
    \draw[draw, thick, |-, blue] (6, 2) -- (8, 2);

    \draw[cr] (2, 1) -- (8.5, 1);
    \draw[draw, thick, blue] (3, 1) -- (8, 1) node[midway] (unbounded) {};

    \draw[dashed, red, thick] (3, -1.5) -- (3, 4);
    \draw[dashed, red, thick] (8, -1.5) -- (8, 4);
    \draw [decorate,decoration={brace, amplitude=5pt, mirror, raise=4ex}] (3, -1) -- (8,-1) node[midway, yshift=-3em]{Timeframe};
    \draw[->] (1.5, -1) -- (9.5, -1);
    \draw [
      postaction={
        draw,
        decoration=ticks,
        segment length=0.5cm,
        decorate,
      }
    ] (2,-1) -- (9,-1);

    \draw (left_bounded.center) to[out=90, in=-180] (8.5, 3) node[right] {left-bounded};
    \draw (right_bounded.center) to[out=90, in=0] (2.5, 3) node[left] {right-bounded};
    \draw (unbounded.center) to[out=-90, in=0] (5, 0) node[left] {unbounded};
    \draw (bounded.center) to[out=90, in=0] (4.5, 4) node[left] {bounded};

  \end{tikzpicture}
  \caption{The impact of the timeframe on data completeness: The concurrent code reviews as communication channels may have started before or ended after the observed timeframe. Due to the cut, the communication channels may cut at their start (right-bound) or at their end (left-bound), or the channel is completely contained (bound) or not contained (unbound) in the timeframe. }
  \label{fig:timeframe}
\end{figure}

Finally, the finite timeframe of our simulation introduces unavoidable edge effects. Information diffusion is a continuous process, but our measurement window imposes artificial start and end points. Some communication channels may begin before or continue after the observation period (\Cref{fig:timeframe}). \Cref{fig:timeframe} shows the distribution of communication channels classified by their temporal boundaries across the studied code review systems: bounded, left-bounded, right-bounded, and unbounded. While the proportions vary between all systems, the share of problematic right-bounded and unbounded channels—where code review participants may be missed and information diffusion potentially underestimated—is consistently small for Android, Visual Studio Code, React, and Spotify (less than about 10\%). In contrast, these edge effects are more pronounced in the code review systems at Microsoft and Trivago, suggesting that we may underestimate the upper bound of information diffusion in those systems. We found no clear explanation for this pattern within the observed timeframe.
\begin{figure*}
  \centering
  \begin{tikzpicture}
    \begin{axis}[
        height=6cm,
        width=\textwidth,
        ybar stacked,
        ymajorgrids,
        ymin=0,
        ymax=1,
        enlarge y limits=0.,
        xtick=data,
        bar width=0.5cm,
        xticklabels={Android, Visual Studio Code, React, Microsoft, Spotify, Trivago},
        xticklabel style={align=center, text width=2cm},
        legend pos=outer north east,
        legend style={
          at={(0.5,1.15)}, 
          anchor=south,    
          legend columns=4, 
          align=left,      
          column sep=1ex,  
          draw=none        
        }
      ]
      \addplot[thin, fill=bounded_color, postaction={}, area legend] table [y=bounded, x expr=\coordindex, col sep=comma] {data/csv/bounds.csv};
      \addlegendentry{bounded}

      \addplot[thin, fill=leftbounded_color, postaction={}, area legend] table [y=left-bounded, x expr=\coordindex, col sep=comma] {data/csv/bounds.csv};
      \addlegendentry{left-bounded}

      \addplot[thin, fill=rightbounded_color, postaction={}, area legend] table [y=right-bounded, x expr=\coordindex, col sep=comma] {data/csv/bounds.csv};
      \addlegendentry{right-bounded}

      \addplot[thin, fill=unbounded_color, postaction={ }, area legend] table [y=unbounded, x expr=\coordindex, col sep=comma] {data/csv/bounds.csv};
      \addlegendentry{unbounded}

    \end{axis}
  \end{tikzpicture}
  \caption{Shares of bounded, left-bounded, right-bounded, and unbounded communication channels among all code review systems within the simulation timeframe. }
  \label{fig:bound}
\end{figure*}
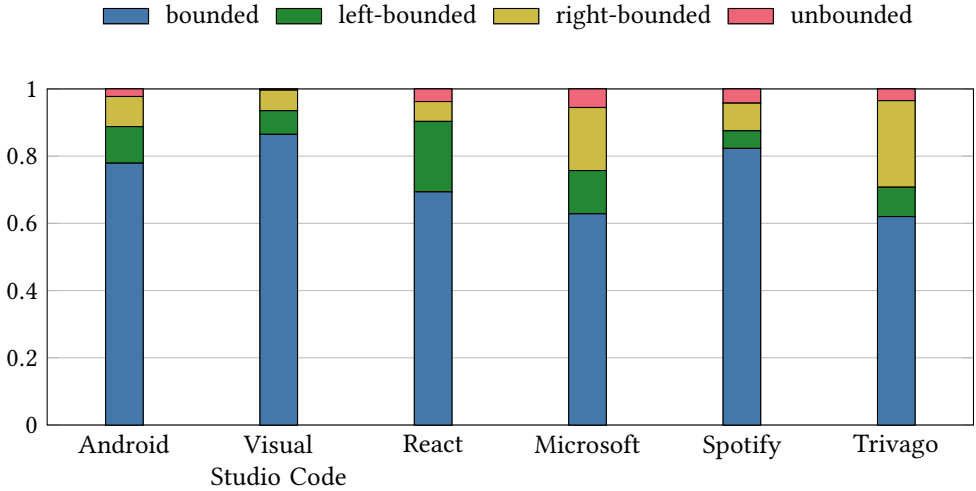

\subsection{Construct Validity}

Our study operationalizes information diffusion using structural metrics derived from simulated code review interactions. These measurements estimate how widely and quickly information could diffuse through code review networks under idealized, best-case assumptions: namely, that information spreads whenever structural and temporal conditions permit. While this approach provides a useful upper bound on diffusion potential, it likely overstates the extent of real-world communication, which depends on human factors such as selective attention, comprehension, memory, and intent. As a result, our structural measurements may partially misalign with the theoretical construct of code review as a human-centered communication network.

In particular, we treat review participation as a proxy for meaningful communication, without accounting for variation in message quality, reviewer influence, or interaction depth. This abstraction overlooks important cognitive and social processes that govern actual information exchange---such as whether reviewers internalize feedback, whether authors incorporate it, and how review discussions are shaped by trust or reputation. These simplifications pose a threat to construct validity, as our metrics may capture structural reach rather than substantive communication.

A further threat arises from the potential inclusion of automated accounts or partially automated review behaviors in our analysis. To preserve the validity of our measurements as proxies for human communication, we aimed to exclude bot activity from our data. This filtering might be more effective in open-source systems, where public metadata and transparent contribution histories enabled fine-grained identification and removal of automated accounts. In closed-source environments, however, access restrictions and anonymized user identifiers limited our ability to reliably detect non-human actors. As a result, instances of developer-driven automation---such as scripted review approvals, preconfigured responses, or notification handlers---may have remained in the dataset. These interactions do not reflect meaningful human engagement and could therefore distort our diffusion measurements. The inclusion of such automation introduces noise into our structural metrics and risks weakening their alignment with the underlying construct of human-to-human information flow in code review.

To improve construct validity in future work, researchers could incorporate behavioral constraints or cognitive models of information processing into simulation logic. For example, empirically derived probabilities of message retention, forwarding, or selective attention---based on factors like review length, comment type, reviewer load, or past collaboration---could bring modeled diffusion closer to real developer behavior. Additionally, refining the operationalization of diffusion through multi-dimensional metrics that integrate structural paths with semantic indicators (e.g., comment sentiment, topic alignment, or question–response dynamics) may yield a richer and more accurate view of how knowledge spreads through code review.

Finally, validating structural diffusion measures against ground truth---such as developer-reported awareness, trace links to bug resolution, or interview data---could help assess whether our metrics truly reflect meaningful communication outcomes. By combining structural models with empirical insights into how developers communicate and learn during code review, future studies can enhance both the realism and theoretical alignment of information diffusion as a construct.

\subsection{External Validity}

As discussed in \Cref{sec:osvscs}, comparing open-source with closed-source software development is inherently difficult since the definition of open source refers only to a specific type of licensing but not to a specific software development process. The open-source projects included in our study---React (maintained by Meta), Android (led by Google), and Visual Studio Code (maintained by Microsoft)---represent a specific subset of the open-source landscape: large, industry-backed initiatives. Although these projects formally meet the definition of open source, they also benefit from substantial institutional support---such as dedicated engineering staff, structured review processes, and sustained corporate oversight. As a result, they occupy a hybrid space that merges elements of community-driven open source with characteristics of enterprise-level software engineering. This hybrid model sets them apart from both traditional closed-source systems and purely community-led open-source projects. While the inclusion of these corporate-backed initiatives provides valuable insight into scalable, production-grade review practices, it also limits the generalizability of our findings to other open-source contexts---particularly those that are grassroots, volunteer-driven, or lacking institutional infrastructure.

Future research could examine whether the communication and diffusion dynamics observed in industry-backed open-source systems extend to smaller,  community-driven open-source projects. These projects may operate under different social norms, resource constraints, and review practices, which could significantly influence how information spreads through their code review networks. Expanding the empirical base to include a broader spectrum of open-source ecosystems is therefore essential to improving the external validity of studies comparing open- and closed-source development.

\subsection{Conclusion Validity}

Like any empirical study, our work is subject to potential research bias---particularly given its nature as a replication study involving an overlap with the original research team. While the experimental design and quantitative approach may reduce certain forms of subjectivity compared to qualitative methods, the interpretation of results still depends on the perspectives and decisions of the researchers involved. This risk is especially relevant in self-replications, where unconscious bias may influence framing, emphasis, or analytical choices. To mitigate this threat, we have made all data and simulation materials publicly available and invite the research community to independently replicate our findings, extend them to other contexts, or challenge our conclusions. In doing so, we hope to contribute to a cumulative body of evidence---a family of replications---that collectively strengthens the empirical foundation of code review as a communication network \cite{Carver2010}.

\section{Conclusion}

Our study provides a first confirmatory perspective on the theory of code review as a communication network. First, we found that code review networks can indeed spread information widely and quickly, supporting the theory of code review as communication acorrding to which information can spread widely and quickly through code review. We found no evidence that larger network size impairs information diffusion. Even within one of the largest code review systems worldwide---Microsoft's internal code review---information still can spread relatively widely and quickly. This suggests that, at least under best-case conditions, code review networks scale without a loss in their capacity to diffuse information. However, this capability is not guaranteed; the ability of code review systems to support information diffusion is not uniformly distributed across our sample. This underscores that effective information diffusion is not an inherent trait of code review, but a property that must be intentionally cultivated through structure, practice, and design.

Second, we found no evidence that larger network size impairs information diffusion. Even within one of the largest code review systems worldwide---Microsoft's internal network---information still can spread relatively widely and quickly. This suggests that, at least under best-case conditions, code review networks scale without a loss in their capacity to diffuse information.

Third, our findings reveal important differences between open-source and closed-source code review systems with regards to the capability of information diffusion. While open-source projects tend to spread information more quickly, they do so across a significantly smaller fraction of participants compared to their closed-source counterparts. This suggests that findings derived from studies of open-source code review may not necessarily generalize to closed-source environments. Given these structural and behavioral differences, we advocate for a critical reevaluation of the generalizations made in the software engineering research community based primarily on open-source code review.

Taken together, our results both reinforce and refine the theory of code review as a communication network, and they highlight important contextual boundaries that future research and practice must take into account.

\section*{Data Availability}
The replication package, including simulation code, datasets, and analysis scripts, is publicly available at
\begin{center}
  \url{https://github.com/michaeldorner/capability-of-code-review-as-communication-network}.
\end{center}

\begin{acks}
  This work was supported by the KKS Foundation through the SERT Project (Research Profile Grant 2018/010) at Blekinge Institute of Technology.
\end{acks}

\printbibliography

\end{document}